\documentclass[12pt]{article}
\pdfoutput=1
\usepackage{color}
\usepackage{epsfig, palatino}
\usepackage{epic}
\usepackage{dsfont}
\usepackage{colonequals}
\usepackage{mathrsfs}
\usepackage{ae} 
\usepackage[T1]{fontenc}
\usepackage[utf8]{inputenc}
\usepackage{amsmath}
\usepackage{amssymb}
\usepackage{graphicx}
\usepackage[normalem]{ulem}
\usepackage{xcolor}
\definecolor{darkblue}{cmyk}{0.9,0.9,0,0}
\usepackage[colorlinks=true,linkcolor=darkblue,citecolor=darkblue,urlcolor=darkblue]{hyperref}
\usepackage{cite}
\usepackage{hyperref}
\usepackage{varioref}
\usepackage{makeidx}
\usepackage[english]{babel}
\usepackage{simplewick}
\usepackage{array}
\usepackage{multirow}
\usepackage[utf8]{inputenc}
\usepackage{empheq}

\newcommand{\jt}{\ensuremath{j_{\tilde{0}}}}

\usepackage{tikzit}
\tikzstyle{point}=[fill=black, draw=black, shape=circle, inner sep=0pt, minimum size=3pt]
\tikzstyle{lr_point}=[fill=red, draw=red, shape=circle, inner sep=0pt, minimum size=2pt]
\tikzstyle{sh_point}=[fill={rgb,255: red,128; green,128; blue,128}, draw={rgb,255: red,128; green,128; blue,128}, shape=circle, inner sep=0pt, minimum size=3pt]
\tikzstyle{sh_lr_point}=[fill={red!40}, draw={red!40}, shape=circle, inner sep=0pt, minimum size=2pt, tikzit fill={rgb,255: red,255; green,128; blue,0}, tikzit draw={rgb,255: red,255; green,128; blue,0}]

\tikzstyle{dashed_st}=[-, dashed]
\tikzstyle{arrow}=[->]
\tikzstyle{dashed_grey}=[-, draw=black, dotted]
\tikzstyle{blue_line}=[draw=blue, ->]
\tikzstyle{energy}=[draw={rgb,255: red,255; green,128; blue,0}, decoration={{snake,amplitude=1pt,segment length=6pt,post length=1pt}}, decorate, ->]
\tikzstyle{blue_line_0}=[-, draw=blue]
\tikzstyle{redline}=[-, draw=red]
\tikzstyle{blue dot}=[scale=0.3, draw=blue, shape=circle, fill=blue]
\tikzstyle{textdot}=[shape=circle, scale=0.7]
\tikzstyle{point}=[fill=black, draw=black, shape=circle, inner sep=0pt, minimum size=3pt]
\tikzstyle{lr_point}=[fill=red, draw=red, shape=circle, inner sep=0pt, minimum size=2pt]
\tikzstyle{sh_point}=[fill={rgb,255: red,128; green,128; blue,128}, draw={rgb,255: red,128; green,128; blue,128}, shape=circle, inner sep=0pt, minimum size=3pt]
\tikzstyle{sh_lr_point}=[fill={red!40}, draw={red!40}, shape=circle, inner sep=0pt, minimum size=2pt, tikzit fill={rgb,255: red,255; green,128; blue,0}, tikzit draw={rgb,255: red,255; green,128; blue,0}]
\tikzstyle{blue_point}=[fill=blue, draw=blue, shape=circle, inner sep=0pt, minimum size=3pt]
\tikzstyle{circle dot}=[fill=white, draw=black, shape=circle, dashed, inner sep=0pt, minimum size=120pt]

\tikzstyle{new edge style 0}=[-, dashed]
\tikzstyle{arrow end}=[->]
\tikzstyle{dashed_st}=[-, dashed]
\tikzstyle{arrow}=[->]
\tikzstyle{dashed_grey}=[-, draw=black, dotted]
\tikzstyle{blue_line}=[draw=blue, ->]
\tikzstyle{energy}=[draw={rgb,255: red,255; green,128; blue,0}, decoration={{snake,amplitude=1pt,segment length=6pt,post length=1pt}}, decorate, ->]
\tikzstyle{blue_line_0}=[-, draw=blue]
\tikzstyle{redline}=[-, draw=red]
 
\usepackage{subcaption}

\usepackage[font={small}]{caption}

\newcommand{\comment}[1]{}

\newcommand{\ee}{\end{align}} 

\newcommand{\eeq}{\end{equation}}

\newcommand{\eeqq}{\end{equation*}}

\newcommand\eeqaa{\end{eqnarray*}}

\newcommand\eeqa{\end{array}}

\newcommand{\eea}{\end{eqnarray}}

\newcommand{\neqa}{\nonumber\end{eqnarray}}

\renewcommand{\d}{\partial}

\newcommand{\<}{{\langle}}
\renewcommand{\>}{{\rangle}}

\newcommand{\re}{\relax{\rm I\kern-.18em R}}

\renewcommand{\sp}{p\hspace{-.40em}/}

\definecolor{darkgreen}{rgb}{0.0, 0.45, 0.0}
\definecolor{mathematicablue}{RGB}{94,130,182}

\def\XXint#1#2#3{{\setbox0=\hbox{$#1{#2#3}{\int}$}
\vcenter{\hbox{$#2#3$}}\kern-.5\wd0}}

\def\su2{{SU(2)}}

\def\[{\left[}
\def\]{\right]}

\def\({\left(}
\def\){\right)}
\def\[{\left[}
\def\]{\right]}

\def\<{\langle}
\def\>{\rangle}

\def\i2{\frac{i}{2}}

\def\O{{\mathcal O}}

\def\spi{\relax{\rm \pi\kern-0.5em /}}
\def\sA{\relax{\rm A\kern-0.5em /}}
\def\sp{\relax{\rm p\kern-0.5em /}}
\def\sd{\relax{\rm \d\kern-0.5em /}}
\def\sk{\relax{\rm k\kern-0.5em /}}
\def\sn{\relax{\rm n\kern-0.5em /}}
\def\sl{\relax{\rm l\kern-0.5em /}}
\def\sP{\relax{\rm P\kern-0.7em /}}
\def\sBethe{\relax{\rm \Bethe\kern-0.5em /}}

\def\2F1{\,_2{\rm F}_1}

\makeindex

\begin{document}

\thispagestyle{empty}

\renewcommand{\thefootnote}{\fnsymbol{footnote}}
\setcounter{page}{1}
\setcounter{footnote}{0}
\setcounter{figure}{0}

\begin{center}
$$$$

{\Large\textbf{\mathversion{bold}
Four point functions in CFT's \\
with slightly broken higher spin symmetry
}\par}
\vspace{1.0cm}

\vspace{1.0cm}

\textrm{ Joao A. Silva}
\\ \vspace{1.2cm}
\footnotesize{\textit{ 
Fields and Strings Laboratory, Institute of Physics, \'Ecole Polytechnique F\'ed\'erale de Lausanne (EPFL)\\
 CH-1015 Lausanne,
Switzerland
%$^\text{\tiny 2}$ CERN, Theoretical Physics Department, 1211 Geneva 23, Switzerland
}  
\vspace{4mm}
}

\par\vspace{1.5cm}

\textbf{Abstract}\vspace{2mm}
\end{center}

We compute spinning four point functions in the quasi-fermionic three dimensional conformal field theory with slightly broken higher spin symmetry at finite t'Hooft coupling. More concretely, we obtain a formula for $\langle j_s  \jt \jt \jt \rangle$, where $j_s$ is a higher spin current and $\jt$ is the scalar single trace operator. Our procedure consists in writing a plausible ansatz in Mellin space and using crossing, pseudo-conservation and Regge boundedness to fix all undetermined coefficients. Our method can potentially be generalised to compute all spinning four point functions in these theories.

%
%To our great surprise we find that a free massive scalar minimally coupled to gravity does not admit UV completion as a consistent theory of quantum gravity within a certain range of masses!

%We compute spinning four point functions in the quasi-fermionic three dimensional conformal field theory with slightly broken higher spin symmetry at finite t'Hooft coupling. More concretely, we conjecture a formula for $\langle j_s  \jt \jt \jt \rangle$, where $j_s$ is a higher spin current and $\jt$ is the scalar single trace operator. Our procedure consists in writing a plausible ansatz in Mellin space and using crossing, pseudo-conservation and softness in the bulk point limit to fix all undetermined coefficients. 

%Mellin space is an important ingredient in our calculation. Our conjecture rests on three technical assumptions. The most important of these is softness in the bulk point limit. We expect our method to generalise for generic spinning four point functions $\langle j_{s_1}  j_{s_2} j_{s_3} j_{s_4}  \rangle$ at finite t'Hooft coupling, namely we guess the form that the corresponding Mellin amplitude should have. The difficulty in implementing an algorithm for generic spinning four point functions lies in the large number of possible conformal structures, but we expect this difficulty to be surmountable by the use of conformal frame techniques.

%

%
%
%
%

%

\noindent

\setcounter{page}{1}
\renewcommand{\thefootnote}{\arabic{footnote}}
\setcounter{footnote}{0}

\setcounter{tocdepth}{2}

 \def\nref#1{{(\ref{#1})}}

\newpage

\tableofcontents

\parskip 5pt plus 1pt   \jot = 1.5ex

\section{Introduction and summary of results}

The dualities between conformal field theories and higher spin gravity theories in AdS are one of the most intriguing topics in the AdS/CFT correspondence. Potentially, these dualities should allow for an improved understanding of the AdS/CFT correspondence, since both sides of the duality are simple, at least when compared to the more standard case of $\mathcal{N}=4$ SYM and type IIB superstring theory\footnote{See \cite{Aharony:2020omh} (which builds on the works \cite{Koch:2010cy, Koch:2014aqa, deMelloKoch:2018ivk}) for recent progress, where the path integral for critical $O(N)$ models was written in terms of higher spin gauge fields defined in the bulk of AdS.}. Of particular interest are CFT's with slightly broken higher spin symmetry, that were studied most notably in the paper by Maldacena and Zhiboedov \cite{Maldacena:2012sf}, where all three point functions of single trace operators at the planar level were computed at finite t'Hooft coupling. In our paper, we compute some four point functions of spinning single trace operators at the planar level at finite t'Hooft coupling. The formulas we obtain are very simple and our formalism, which is based on pure CFT arguments in which Mellin space plays an important role, potentially paves the way for the computation of all spinning four point functions.

CFT's with slightly broken higher spin symmetry are large $N$ CFT's where higher spin symmetry is broken by $1/N$ effects. There are two such theories, the quasi-boson theory and the quasi-fermion theory, which are defined in $3$ dimensions. We will focus on the quasi-fermion theory. This theory depends on two parameters, $\tilde{N}$ and $\tilde{\lambda}$ (we follow the notation of \cite{Maldacena:2012sf}). We will study the theory at the planar level, i. e. at leading order in $\tilde{N}$. In that case the theory interpolates between the free fermion theory at $\tilde{\lambda}=0$ and the critical point of the $O(N)$ model (critical boson) at $\tilde{\lambda}=\infty$.

Being a large $N$ theory, the spectrum of the quasi-fermion theory organises into single and multitrace primary operators. Let us describe the single trace operators. There is one single trace operator for each even spin $s=0, 2, ...$. The scalar primary, which we will denote by $\jt$, has dimension $2+O(\frac{1}{\tilde{N}})$ \cite{Jain:2019fja}. The spin $2$ primary $j_2$ is exactly conserved. A higher spin primary $j_s$ of spin $s>2$ has dimension $s+1$ and acquires anomalous dimensions of $O(\frac{1}{\tilde{N}})$ \cite{Giombi:2016zwa}, \cite{Giombi:2017rhm}.

This theory is believed to be solvable in the planar limit. In \cite{Maldacena:2012sf} three point functions of single trace operators were computed at the planar level and for finite $\tilde{\lambda}$ through the use of slightly broken higher spin Ward identities\footnote{This calculation was reproduced using higher spin techniques in \cite{Skvortsov:2018uru}, where also the parity odd structures were given.}. In \cite{Turiaci:2018dht} four point functions of scalar operators were computed using the Lorentzian inversion formula and Schwinger-Dyson equations. In \cite{Li:2019twz} the four point function $\langle j_2 \jt \jt \jt\rangle$ was computed using the pseudo-conservation equations\footnote{Correlators in ABJ theory were computed using slightly broken higher spin symmetry in \cite{Binder:2021cif}.}. 

We obtain a formula for $\langle j_s \jt \jt \jt\rangle$ for generic spin $s \geq 4$:
\begin{align}\label{n and l intro}
\langle j_s \jt \jt \jt \rangle =\frac{1}{\tilde{N} \sqrt{1+\tilde{\lambda}^2}} \langle j_s \jt \jt \jt \rangle_{ff}
+ \frac{\tilde{\lambda}}{\tilde{N} \sqrt{1+\tilde{\lambda}^2}} \langle j_s \jt \jt \jt \rangle_{cb} ,
\end{align}
where $\langle j_s \jt \jt \jt \rangle_{ff}$ is the correlator in the free fermion theory (which is fully known) and $\langle j_s \jt \jt \jt \rangle_{cb}$ is the corresponding correlator in the critical boson theory. The critical boson theory is the IR fixed point of the theory of $\tilde{N}$ free real scalar fields perturbed by $(\phi_i \phi_i)^2$. 

This result agrees with the 3d bosonization picture advanced in \cite{Maldacena:2012sf}, where it is proposed that the quasi-fermionic theory interpolates between a tridimensional theory of $\tilde{N}$ free fermions and the critical theory of $\tilde{N}$ bosons, in the limits $\tilde{\lambda} \rightarrow 0$ and $\tilde{\lambda} \rightarrow \infty$ respectively.

We obtain that
\begin{align}\label{Mellin s 2 intro}
\langle j_s \jt \jt \jt \rangle_{cb}  =  |x_1-x_3|^{-4 s-2}  |x_2-x_3|^{2s-1} 
 |x_2-x_4|^{-2s-3}  |x_3-x_4|^{2 s-1} \\
\times \sum_{k=0}^s \int\int \frac{d \gamma_{12} d{\gamma_{14}}}{(2 \pi i)^2}  M(\gamma_{12}, \gamma_{14}; s, k) u^{- \gamma_{12}} v^{-\gamma_{14}} V(1; 2, 3)^{s-k} V(1; 3, 4)^k, \nonumber
\end{align}
where $V(i; j, k)$ is a conformal structure (see (\ref{struct V})) and $u$ and $v$ are the usual conformal cross ratios. $M(\gamma_{12}, \gamma_{14}; s, k)$ is equal to
\begin{align}\label{ansatz Mellin s000 intro}
M(\gamma_{12}, \gamma_{14}; s, k)= \Gamma (-k+\gamma_{14}-1) \Gamma\left(-k+\gamma_{14}+\frac{1}{2}\right) \Gamma(s - \gamma_{12} - \gamma_{14}) \\
\times \Gamma\left(s - \gamma_{12} -\gamma_{14} +\frac{3}{2} \right) \Gamma(k-s+\gamma_{12}-1) \Gamma\left(k-s+\gamma_{12}+\frac{1}{2}\right) p(\gamma_{12}, \gamma_{14}; s, k), \nonumber
\end{align}
where $p(\gamma_{12}, \gamma_{14}; s, k)$ is a polynomial in $\gamma_{12}$ and $\gamma_{14}$. This polynomial is fully determined by crossing, pseudo-conservation and Regge boundedness, see equations (\ref{crossing1}) and (\ref{crossing2}), see (\ref{pseudoconservation polynomial}) and see also (\ref{bound on chaos polynomial 1}), (\ref{bound on chaos polynomial 2}) and  (\ref{bound on chaos polynomial 3}).

%In section \ref{algorithm in Mellin space} we show how formula (\ref{ansatz Mellin s000 intro}) solves the crossing and pseudo-conservation equations. In section \ref{bound chaos section} we show that formula (\ref{ansatz Mellin s000 intro}) is the unique solution to the conservation equations consistent with the bound on chaos. One could think that  

We explain in section \ref{algorithm in Mellin space} how formula (\ref{n and l intro}) solves the crossing and pseudo-conservation equations and correctly accounts for the exchange of single trace operators with the OPE coefficients derived in \cite{Maldacena:2012sf}. In section \ref{bound chaos section} we show that formula (\ref{n and l intro}) is the unique solution to the pseudo-conservation and crossing equations which is consistent with the bound on chaos. In particular we analyse AdS contact diagrams for $\langle j_s \jt \jt \jt \rangle$ and we conclude that such diagrams violate the bound on chaos, provided $s\geq 4$. In section \ref{OpenDirections} we discuss open directions. In appendix \ref{app: bulk} we study the bulk point limit of $\langle j_s \jt \jt \jt\rangle$. In appendix \ref{app:algorithm in position space} we calculate $\langle j_s \jt \jt \jt\rangle$ in position space for spins $s=2, ..., 14$. This calculation agrees with the Mellin space result. In appendix \ref{app: mixed Fourier} we recompute $\langle j_2 \jt \jt \jt\rangle$ by solving the higher spin Ward identities.

\section{The bootstrap of $\langle j_s \jt \jt \jt \rangle$}\label{algorithm in Mellin space}

%\textbf{$\tilde{\lambda}$ and $\tilde{N}$ dependence}. 

We will compute $\langle j_s \jt \jt \jt \rangle$. Let us start by examining the $\tilde{N}$ and $\tilde{\lambda}$ dependence. It is expected that the quasi-fermion theory interpolates between a theory of $\tilde{N}$ free fermions at $\tilde{\lambda}=0$ and the critical boson theory at $\tilde{\lambda} = \infty$. 

%The critical boson theory is the IR fixed point of the theory of $\tilde{N}$ free real scalar fields perturbed by $(\phi_i \phi_i)^2$. 

We will work in a normalization where $\langle j_s j_s \rangle \sim 1$, i.e. two point functions of single trace operators do not depend on $\tilde{N}$ or $\tilde{\lambda}$. We use the $\sim$ sign to mean that we do not keep track of numerical factors, but we do keep track of the $\tilde{N}$ and $\tilde{\lambda}$ dependence. Thus, $\langle j_s \jt \jt \jt \rangle \sim \frac{1}{\tilde{N}}$. At this order, we can only have exchanges of single trace operators or double trace operators $[\jt, \jt]$ or $[j_s, \jt]$.

%\footnote{To the best of our knowledge, three point functions of two single trace operators and one double trace operator at $O(\frac{1}{\tilde{N}})$ have not yet been written in the literature.}.

%where by $[j_{s_1}, j_{s_2}]$ we mean a generic double trace operator made of $j_{s_1}$ and $j_{s_2}$

Let us consider exchanges of single trace operators. The relevant three point functions are $\langle j_s \jt j_{s'} \rangle$ and $\langle j_{s'} \jt \jt \rangle$, with $s' \geq 2$. Note that $\langle \jt \jt \jt \rangle = 0$ \cite{Maldacena:2012sf}. From \cite{Maldacena:2012sf} we see that $\langle j_s \jt \jt \rangle \sim \frac{1}{\sqrt{\tilde{N}}}$. There are two possible structures for $\langle j_s \jt j_{s'} \rangle$, the fermion and the odd structure. We have that $\langle j_s \jt j_{s'} \rangle_{fermion} \sim \frac{1}{\sqrt{\tilde{N}} \sqrt{1 + \tilde{\lambda}^2} }$ and $\langle j_s \jt j_{s'} \rangle_{odd} \sim \frac{\tilde{\lambda}}{\sqrt{\tilde{N}} \sqrt{1 + \tilde{\lambda}^2} }$.

Based on this we propose the following ansatz
\begin{align}\label{n and l}
\langle j_s \jt \jt \jt \rangle =\frac{1}{\tilde{N} \sqrt{1+\tilde{\lambda}^2}} \langle j_s \jt \jt \jt \rangle_{ff} 
+ \frac{\tilde{\lambda}}{\tilde{N} \sqrt{1+\tilde{\lambda}^2}} \langle j_s \jt \jt \jt \rangle_{cb} ,
\end{align}
where $\langle j_s \jt \jt \jt \rangle_{ff}$ is the four point function in the free fermion theory, whose form can be read in \cite{Didenko:2013bj}. To the best of our knowledge, $\langle j_s \jt \jt \jt \rangle_{cb}$ has not yet been computed and it will be the subject of this section to do precisely that. We attached the subscript $_{cb}$ since it is expected that it corresponds to a four point function in the critical boson theory.

The reader might be confused about the factor of $\frac{1}{\tilde{N}}$. In our normalization, $\lim_{\tilde{\lambda} \rightarrow 0} \langle j_s \jt \jt \jt \rangle = \frac{\langle j_s \jt \jt \jt \rangle_{ff} }{\tilde{N}} $ and  $\lim_{\tilde{\lambda} \rightarrow \infty} \langle j_s \jt \jt \jt \rangle = \frac{\langle j_s \jt \jt \jt \rangle_{cb} }{\tilde{N}} $. Given that the quasi-fermionic theory interpolates between a theory of $\tilde{N}$ free fermions and the critical theory of $\tilde{N}$ bosons, the reader might be confused about why there is a factor of $\frac{1}{\tilde{N}}$. There are two things happening in this context. First, when we write $\langle j_s \jt \jt \jt \rangle_{ff}$ and $\langle j_s \jt \jt \jt \rangle_{cb}$ we have decided to factor out the dependence on $\tilde{N}$. Second, the reader might thus have expected to encounter $\langle j_s \jt \jt \jt \rangle \sim \tilde{N}$, but this is only true when two point functions are normalized such that $\langle j_s j_s \rangle \sim \langle \jt \jt \rangle \sim \tilde{N}$, whereas we are using different normalizations, namely $\langle j_s j_s \rangle \sim \langle \jt \jt \rangle \sim 1$. This justifies why does $\langle j_s \jt \jt \jt \rangle \sim \frac{1}{\tilde{N}}$.

%We will compute $\langle j_s \jt \jt \jt \rangle_{cb}$ using a conformal bootstrap method. Let us briefly mention three other techniques that in principle determine $\langle j_s \jt \jt \jt \rangle_{cb}$. These can be used in the future to check our calculations. One technique consists in computing correlators in the critical boson theory from integrals of correlators in free theory \cite{Giombi:2017mxl}, \cite{Giombi:2018vtc}. Another possibility is to use Schwinger-Dyson equations \cite{Giombi:2011kc}, \cite{Bedhotiya:2015uga}. A third possibility is to use slightly broken higher spin Ward identities. Indeed, in the appendix to this paper we compute $\langle j_2 \jt \jt \jt  \rangle_{cb}$ from such identities using a mixed Fourier transform\footnote{The result for $\langle j_2 \jt \jt \jt  \rangle_{cb}$ was published first by \cite{Li:2019twz}. The result in \cite{Li:2019twz} matches the computation that we performed independently.}. This is a check on our main method that we will now describe.  

We can write parity even and parity odd structures for the correlator $\langle j_s \jt \jt \jt \rangle$. The parity odd structures are realised in the free fermion theory. This is because $\jt$ is parity odd in the free fermion theory. The parity even structures are realised in the quasi-boson theory. Thus, we write
\begin{align}\label{corr with struct}
\langle j_s \jt \jt \jt \rangle_{cb} = \sum_{k=0}^s f_k(x_{ij}) V(1; 2, 3)^{s-k} V(1; 3, 4)^k,
\end{align}
where $V(i; j, k)$ is a conformal structure which is given in embedding space \cite{Costa:2011mg} by
\begin{align}\label{struct V}
V(i; j, k) = \frac{(Z_i \cdot P_j) (P_i \cdot P_k) - (Z_i \cdot P_k) (P_i \cdot P_j) }{P_j \cdot P_k}.
\end{align}
$P_i$ and $Z_i$ are null vectors on $\mathbb{R}^{3,2}$. $Z_i$ encodes the spinning indices. $f_k(x_{ij})$ is a function of the distances between the points, with appropriate weights on each of the points. We find it advantageous to consider the Mellin representation 
 \begin{align}\label{def f}
f_k(x_{ij})=\int [\frac{d \gamma_{ij}}{2 \pi i}] \hat{M} (\gamma_{ij}; s, k) x_{ij}^{-2 \gamma_{ij}}, \\
\sum_{j \neq 1} \gamma_{1j} = 2 s + 1, ~~ \sum_{j \neq i} \gamma_{ij} = 2, ~~ i=2,3,4. \nonumber
 \end{align}

%\begin{align}\label{Mellin spin s}
%\langle j_s \jt \jt \jt \rangle = \sum_{k=0}^s \prod_{i<j} \int [\frac{d \gamma_{ij}}{2 \pi i}]M(\gamma_{ij}; s, k) x_{ij}^{-2 \gamma_{ij}} w(1; 2, 3)^{s-k} w(1; 3, 4)^k,
%\end{align}
%\begin{align}
%\sum_{j \neq 1} \gamma_{1j} = 2 s + 1, ~~ \sum_{j \neq i} \gamma_{ij} = 2, ~~ i=2,3,4. \nonumber
%\end{align}
(\ref{corr with struct}) can be rewritten as
\begin{align} \label{Mellin s 2}
\langle j_s \jt \jt \jt \rangle_{cb}  = |x_1-x_3|^{-4 s-2}  |x_2-x_3|^{2s-1} 
 |x_2-x_4|^{-2s-3}  |x_3-x_4|^{2 s-1}  \\
\times \sum_{k=0}^s \int\int \frac{d \gamma_{12} d{\gamma_{14}}}{(2 \pi i)^2}  \hat{M}(\gamma_{12}, \gamma_{14}; s, k) u^{- \gamma_{12}} v^{-\gamma_{14}} V(1; 2, 3)^{s-k} V(1; 3, 4)^k.  \nonumber
\end{align}
We will call $\hat{M}(\gamma_{12}, \gamma_{14}; s, k)$ the Mellin amplitude\footnote{Spinning Mellin amplitudes are analysed in \cite{Chen:2017xdz, Goncalves:2014rfa, Faller:2017hyt}. The definitions slightly differ among these works, but at least concerning conformal four point functions the basic idea is to decompose the correlator in a basis of spinning structures and take the Mellin transform with respect to each function of the positions multiplying each structure. Up to now all works use the embedding space formalism, which has the serious drawback of involving many degeneracies for arbitrary spinning correlators. For generic spinning correlators, we think it would be interesting to define Mellin amplitudes with the conformal frame formalism \cite{Kravchuk:2016qvl, Cuomo:2017wme}, which does not have the problem of degeneracies. We think that it is an interesting problem to work out the poles and residues of the Mellin amplitude for spinning correlators using the conformal frame formalism.}. 

The location of the poles of the Mellin amplitude is related to the operator product expansion of the external operators. Let us make this point explicitly. Consider two external operators $O_1$, $O_2$ of dimensions $\Delta_1$, $\Delta_2$ and spins $s_1$, $s_2$ and suppose they exchange an operator of dimension $\Delta$ and spin $s$. Then the most singular term in the lightcone OPE is 
\begin{align}
\O_{\mu_1 ... \mu_{s_1}}(x) \O_{\nu_1 ... \nu_{s_2}}(0) \supset \frac{\O_{\rho_1 ... \rho_s}(0) x^{\rho_1} ... x^{\rho_s} }{(x^2)^{\frac{\Delta_1 + \Delta_2 + s_1 + s_2}{2} - \frac{\tau}{2}} } x_{ \{ \mu_1 ... \mu_{s_1} \} } x_{ \{  \nu_1 ... \nu_{s_2} \} } \big(1 +  O(x^2) \big),
\end{align}
where $\tau=\Delta-s$. From this logic we expect the Mellin amplitude to have poles at $\gamma_{12} = \frac{\Delta_1 + \Delta_2 + s_1 + s_2}{2} - \frac{\tau}{2} - n$, where $n$ is a nonnegative integer. 

%In order to predict the poles of the Mellin amplitude given knowledge of the twists of the exchanged operators, we need to consider the poles in every variable $$

For $\langle j_s \jt \jt \jt \rangle$ all OPE channels are equal. To order $\frac{1}{\tilde{N}}$ there can be exchanges of higher spin currents and double traces $[j_s, \jt ]$ and $[\jt, \jt]$, which have twist $1$, $3$ and $4$ respectively. This motivates the following ansatz
\begin{align}\label{ansatz Mellin s000}
\hat{M}(\gamma_{12}, \gamma_{14}; s, k)= \Gamma (-k+\gamma_{14}-1) \Gamma\left(-k+\gamma_{14}+\frac{1}{2}\right) \Gamma(-s+\gamma_{13}-1) \\
\times \Gamma\left(-s+\gamma_{13}+\frac{1}{2}\right) \Gamma(k-s+\gamma_{12}-1) \Gamma\left(k-s+\gamma_{12}+\frac{1}{2}\right) p(\gamma_{12}, \gamma_{14}; s, k), \nonumber
\end{align}
where $\gamma_{13} =  2 s + 1 - \gamma_{12} - \gamma_{14}$. The $\Gamma$ functions contain all the poles implied by the OPE. For this reason we assume that $p(\gamma_{12}, \gamma_{14}; s, k)$ is a polynomial in the Mellin variables.

The bound on chaos \cite{Maldacena:2015iua} bounds the degree of the polynomial $p(\gamma_{12}, \gamma_{14}; s, k)$. This is worked out in section (\ref{bound chaos section}), see  (\ref{bound on chaos polynomial 1}), (\ref{bound on chaos polynomial 2}) and (\ref{bound on chaos polynomial 3}) for the precise formulas. Furthermore, $\langle j_s \jt \jt \jt \rangle$ is constrained by invariance under interchange of points $2 \leftrightarrow 3$ and $2 \leftrightarrow 4$. This crossing symmetry implies the equations 
\begin{align}
p(\gamma_{12}, \gamma_{14}; s, k)=
   \sum _{k_2=k}^s (-1)^{k_2} \binom{k_2}{k} p(2s + 1-k_2 - \gamma_{12} - \gamma_{14},\gamma_{14}-k+k_2; s, k_2) \label{crossing1} , \\
p(\gamma_{12}, \gamma_{14}; s, k)= p(\gamma_{14}, \gamma_{12}; s, s-k)  . \label{crossing2} 
\end{align}

$\langle j_s \jt \jt \jt \rangle$ is constrained by pseudoconservation of $j_s$. We implement this condition in embedding space. The differential operator for conservation is $\frac{\partial}{\partial P_1^A} D_A$, where
\begin{align}\label{conservation operator}
D_A= ( \frac{d}{2}-1+  Z_1 \cdot \frac{\partial}{\partial Z_1})\frac{\partial}{\partial Z_1^A} - \frac{1}{2} (Z_1)_A \frac{\partial^2}{\partial Z_1 \cdot \partial Z_1}.
\end{align}

Since $\partial \cdot j_s$ is a primary operator of spin $s-1$ and dimension $s+2$, then $\langle \partial \cdot j_s \jt \jt \jt \rangle$ is a conformal four point function of primary operators. $\langle \partial \cdot j_s \jt \jt \jt \rangle$ factorizes into products of a two point function times a three point function. Such a four point function is made up of powers of $u$ and of $v$ and so its Mellin amplitude vanishes.

%Such a four point function is very simple and so we strongly expect it to have vanishing Mellin amplitude. 

Four point functions of scalars with vanishing Mellin amplitudes were analysed in \cite{Penedones:2019tng}, see in particular section E.E.1. A similar analysis can be performed for the spinning case, though we will not pursue it here. The important conclusion is that in Mellin space pseudoconservation is the same as conservation. In other words, $\langle \partial \cdot j_s \jt \jt \jt \rangle$ has a vanishing Mellin amplitude.

Pseudoconservation implies the equation
\begin{align}\label{pseudoconservation polynomial}
\sum_{i_1=-1}^1 \sum_{i_2=-1}^1 \sum_{i_3=-1}^2 a_{i_1, i_2, i_3} (\gamma_{12}, \gamma_{14}) p(\gamma_{12}+i_1, \gamma_{14}+i_2; s, k+i_3) =0.
\end{align}
%\begin{align}
%a_1( \gamma_{12}, \gamma_{14}; s, k)   p(1+\gamma_{12}, -1+\gamma_{14}; s, -1+k) +  a_2( \gamma_{12}, \gamma_{14}; s, k) p(1+\gamma_{12}, -1+\gamma_{14}; s, k) \\
%\nonumber +  a_3( \gamma_{12}, \gamma_{14}; s, k) p(\gamma_{12}, \gamma_{14}; s, k) + a_4( \gamma_{12},-1+\gamma_{14}; s, k) p(\gamma_{12}, \gamma_{14}; s, k) \\
%\end{align}
%\begin{align}
%-\frac{1}{2} (\gamma_{12}-1) \left(2 k^2+3 k+1\right) \left(2 \gamma_{12}^2+\gamma_{12} (4 \gamma_{14}-4 s-3)+2 \gamma_{14}^2-\gamma_{14} (4 s+3)+s (2 s+3)\right) \\
%\times  p(-1+\gamma_{12}, \gamma_{14}; s, 1+k) 
%+\frac{1}{2} (\gamma_{12}-1) \left(2 k^2+3 k+1\right) \left(2 \gamma_{14}^2-\gamma_{14} (4 k+5)+2 k^2+5 k+2\right) \nonumber \\
%\times  p(-1+\gamma_{12},1+\gamma_{14}; s, 1+k) +(\gamma_{12}-1) \left(k^2+3 k+2\right) \nonumber \\ \times  \left(2 \gamma_{12}^2+\gamma_{12} (4 k-4 s-1)+2 k^2-k (4 s+1)+2 s^2+s-1\right)  p(-1+\gamma_{12},1+\gamma_{14}; s, 2+k) \nonumber \\
%+\frac{1}{2} \left(2 k^2-4 k s+k+s (2 s-1)\right) \left(2 \gamma_{12}^2+\gamma_{12} (4 \gamma_{14}-4 s-3)+2 \gamma_{14}^2-\gamma_{14} (4 s+3)+s (2 s+3)\right) \nonumber \\
%\times (\gamma_{14}-1) p(\gamma_{12},-1+\gamma_{14}; s, k)-\frac{1}{2} \left(2 \gamma_{14}^2-\gamma_{14}(4 k+5)+2 k^2+5 k+2\right) (k-s)  \nonumber \\
%\times (-2 \gamma_{12}(k+s)+\gamma_{14} (2 k-2 s+1)+s (2 s+1) )p(\gamma_{12},\gamma_{14}; s, k)\nonumber
%\end{align}
The coefficients are written in the appendix \ref{app:dump}, see formula (\ref{pol pseudo coefs}).

The crossing equations (\ref{crossing1}), (\ref{crossing2}), the pseudoconservation equation (\ref{pseudoconservation polynomial}) and Regge boundedness (\ref{bound on chaos polynomial 1}), (\ref{bound on chaos polynomial 2}) and (\ref{bound on chaos polynomial 3}) determine $p(\gamma_{12}, \gamma_{14}; s, k)$ up to a multiplicative constant. This has to do with the fact that we have not picked a normalization for the higher spin current $j_s$. It is simple to solve this set of equations in a computer algebra system for each spin $s$. We find that the solution always has the form
\begin{align}
p(\gamma_{12}, \gamma_{14}; s, k)= \sum_{k_1=0}^k \sum _{k_2=0}^{s-k} b(s,k; k_1, k_2) \gamma_{12}^{k_2} \gamma_{14}^{k_1}, ~~~ k \leq \frac{s}{2} \\
p(\gamma_{12}, \gamma_{14}; s, k)= p(\gamma_{14}, \gamma_{12}; s, s- k), ~~~ k > \frac{s}{2} .
\end{align}

$p(\gamma_{12}, \gamma_{14}; s, k)$ turns out to have degree $s$. Using a laptop we generated solutions up to spin $40$. Picking a normalization in which $p(\gamma_{12}, \gamma_{14}; s, k=0) \supset 1$, we find as an example that for $s=4$ we have
\begin{align}
p(\gamma_{12}, \gamma_{14}; s=4, k=0) = 1 -\frac{19 \gamma_{12}}{20}+\frac{119 \gamma_{12}^2}{360}-\frac{\gamma_{12}^3}{20}+\frac{\gamma_{12}^4}{360}, \\
p(\gamma_{12}, \gamma_{14}; s=4, k=1) =-\frac{8}{15} +\frac{4 \gamma_{12}}{9}-\frac{11 \gamma_{12}^2}{90} + \frac{\gamma_{12}^3}{90}+\left(\frac{2}{5}-\frac{11 \gamma_{12}}{30} +\frac{\gamma_{12}^2}{9}-\frac{\gamma_{12}^3}{90} \right ) \gamma_{14} ,\nonumber \\
p(\gamma_{12}, \gamma_{14}; s=4, k=2) = \frac{1}{5} -\frac{4 \gamma_{12}}{15} +\frac{\gamma_{12}^2}{15}+\left(-\frac{4}{15} +\frac{11 \gamma_{12}}{36} -\frac{13 \gamma_{12}^2}{180} \right)\gamma_{14} \nonumber \\
+ (\frac{1}{15} -\frac{13 \gamma_{12}}{180} +\frac{\gamma_{12}^2}{60})\gamma_{14}^2,  \nonumber \\
p(\gamma_{12}, \gamma_{14}; s=4, k=3) =-\frac{8}{15} +\frac{4 \gamma_{14}}{9}-\frac{11 \gamma_{14}^2}{90} + \frac{\gamma_{14}^3}{90}+\left(\frac{2}{5}-\frac{11 \gamma_{14}}{30} +\frac{\gamma_{14}^2}{9}-\frac{\gamma_{14}^3}{90} \right ) \gamma_{12}, \nonumber \\
p(\gamma_{12}, \gamma_{14}; s=4, k=4) = 1 -\frac{19 \gamma_{14}}{20}+\frac{119 \gamma_{14}^2}{360}-\frac{\gamma_{14}^3}{20}+\frac{\gamma_{14}^4}{360}. \nonumber
\end{align}
We write this correlator in position space in appendix \ref{app:dump}, see formula (\ref{spin 4 position}).

In appendix \ref{app:algorithm in position space} we implement an algorithm to compute $\langle j_s \jt \jt \jt \rangle$ in position space. We managed to determine $\langle j_s \jt \jt \jt \rangle$ in position space for spins $2 ,..., 14$ using this algorithm. We write the formulas for the correlators in position space in an accompanying notebook. Taking the Mellin transform we get precisely the same as we get with the procedure in Mellin space. The advantage of Mellin space is that it allows to write equations (\ref{crossing1}), (\ref{crossing2}) and (\ref{pseudoconservation polynomial}) that determine the solution for generic $s$.

%\footnote{One might also worry about the fact that pseudo-conservation is like conservation in Mellin space, so perhaps we are losing information. This can be explained in the following way. When we act on (\ref{Mellin s 2}) with the conservation operator, we generate many terms and we need to gather them into conformal structures for $\langle \partial \cdot j_s \jt \jt \jt \rangle$. In practice, we gather many integrals and we need to be careful with the integration contour. When we bring all terms into the same integration contour, we might cross poles that generate extra terms. These extra terms should be very simple and precisely equal to the product of two point functions times three point functions that constitute $\langle \partial \cdot j_s \jt \jt \jt \rangle$. In practice, this is cumbersome and we did not carry this out because we can check the Mellin calculation with the position space calculation.}.  

Let us mention some checks on our solution. One such check is compatibility of the pseudo-conservation equations with conformal symmetry. $\partial  \cdot j_s$ is a conformal primary at leading order in $\frac{1}{\tilde{N}}$. $\partial \cdot j_s$ can have contributions coming from $[j_{s_1}, \jt ]$ and $[j_{s_1}, j_{s_2} ]$. Only the former matter since we are interested in $\langle j_s \jt \jt \jt \rangle$. More precisely,
\begin{align}
\partial \cdot j_s \supset \sum_{s_1=2}^{s-2} \sum_{m=0}^{s-s_1-1} c_m \partial^m j_{s_1} \partial^{s-s_1-1-m} \jt.
\end{align}
The coefficients $c_m$ are fixed by conformal symmetry (see formula (\ref{primary check})). When we run our algorithm in position space we do not need to input the values of $c_m$, we prefer to keep them unknown. It turns out that our algorithm fixes $c_m$ in agreement with (\ref{primary check}). This is an important check on our results. 

%\JS{actually, run the notebook and check it again}

We also checked that the short distance limit of our expression for $\langle j_s \jt \jt \jt \rangle_{cb}$ agrees with the correct three point structures for the exchange of higher spin currents. Let us take $s=4$ for concreteness. The short distance limit $u \rightarrow 0$ captures the exchange of the higher spin currents in the s-channel. If afterwards we take $v \rightarrow 1$, we find that the correlator behaves as 
\begin{align}\label{euclidean OPE check}
\lim_{v \rightarrow 1} \lim_{u \rightarrow 0} \langle j_s \jt \jt \jt \rangle_{cb} \sim \sum_{J=2}^{\infty} \frac{1}{u^5} \frac{x_{34}^7   }{x_{13}^7 x_{14}^{11} x_{23}^4} (1-v)^J V(1; 2, 3)^4
\end{align}
The $\sim$ sign means that we just keep track of the conformal structure that appears, but we do not keep track of numerical coefficients. (\ref{euclidean OPE check}) is matched by the behaviour of conformal blocks of higher spin currents in the same limit.

Formula (\ref{n and l intro}) correctly accounts for the exchange of single trace operators in $\langle j_s \jt \jt \jt \rangle$. However, it is not obvious that it correctly accounts for the exchange of double trace operators. Indeed, one can imagine adding to (\ref{n and l intro}) AdS contact diagrams, which are solutions to crossing that only involve the exchange of double trace operators. By taking linear combinations of AdS contact diagrams one can furthermore obtain solutions to the conservation equations. However, in the next section we consider such linear combinations and show that they always violate the bound on chaos. For this reason, it is not legal to add them to (\ref{n and l intro}).

\section{Bound on chaos for $\langle j_s \jt \jt \jt \rangle$} \label{bound chaos section}

%(\ref{recap bound on chaos})

The bound on chaos \cite{Maldacena:2015waa} constrains the Regge limit of $\langle j_s \jt \jt \jt \rangle$. In this section we review the bound on chaos and derive its consequences for $\langle j_s \jt \jt \jt \rangle$. There are two possible structures one can write for $\langle j_s \jt \jt \jt \rangle$. One structure involves the $\epsilon$ tensor and the other one does not. We examine the two cases separately in sections (\ref{bound on chaos parity even}) and (\ref{bound on chaos parity odd}) and derive bounds on the Regge growth of the Mellin amplitude for both of these cases. 

Solutions to crossing that only involve the exchange of double twist operators are given by AdS contact diagrams. This was proven in \cite{Heemskerk:2009pn}, for the special case of four point functions of external scalars. We will assume that such a result holds for any n-point function of spinning conformal primaries. We study AdS contact diagrams in sections (\ref{Regge limit parity even}) and (\ref{bound on chaos parity odd}). Our main conclusion is that AdS contact diagrams for $\langle j_s \jt \jt \jt \rangle$ are incompatible with the bound on chaos, provided $s \geq 4$. For $s=2$ we construct the contact diagrams that are compatible with the bound on chaos, see formulas (\ref{contact diagram spin 2 even}) and (\ref{contact diagram spin 2 odd}). This completes the proof of formula (\ref{n and l intro}).

%\JS{We need to consider also the $\epsilon$ structure for $\langle j_s \jt \jt \jt \rangle$ and demonstrate that contact AdS diagrams violate the bound on chaos. For this we need to to: write down the structures (a), derive the chaos bound (b), write down the Mellin amplitude of contact AdS diagrams (c), impose crossing and conservation and observe that such solutions violate the bound on chaos (d). Notice there is no need to explicitly construct AdS contact diagrams. What we need are the analogues of equations: (a) (\ref{Mellin Regge appendix 1}), (\ref{bound on chaos polynomial 1}) (b), (\ref{ansatz Contact Diagrams s000}) (c) . This should take a day of work.}

\subsection{Review of the bound on chaos and Rindler positivity}\label{recap bound on chaos}

Conformal field theories are constrained by the Regge behaviour of Lorentzian correlators. For nonperturbative CFT's, correlators in the Regge limit are bounded by the Euclidean OPE in the first sheet. For large N CFT's one needs to use the bound on chaos to bound correlators in the Regge limit. In this subsection we review the bound on chaos \cite{Maldacena:2015waa}.

\begin{figure}[h]
  \centering
  \includegraphics[width=0.5\textwidth]{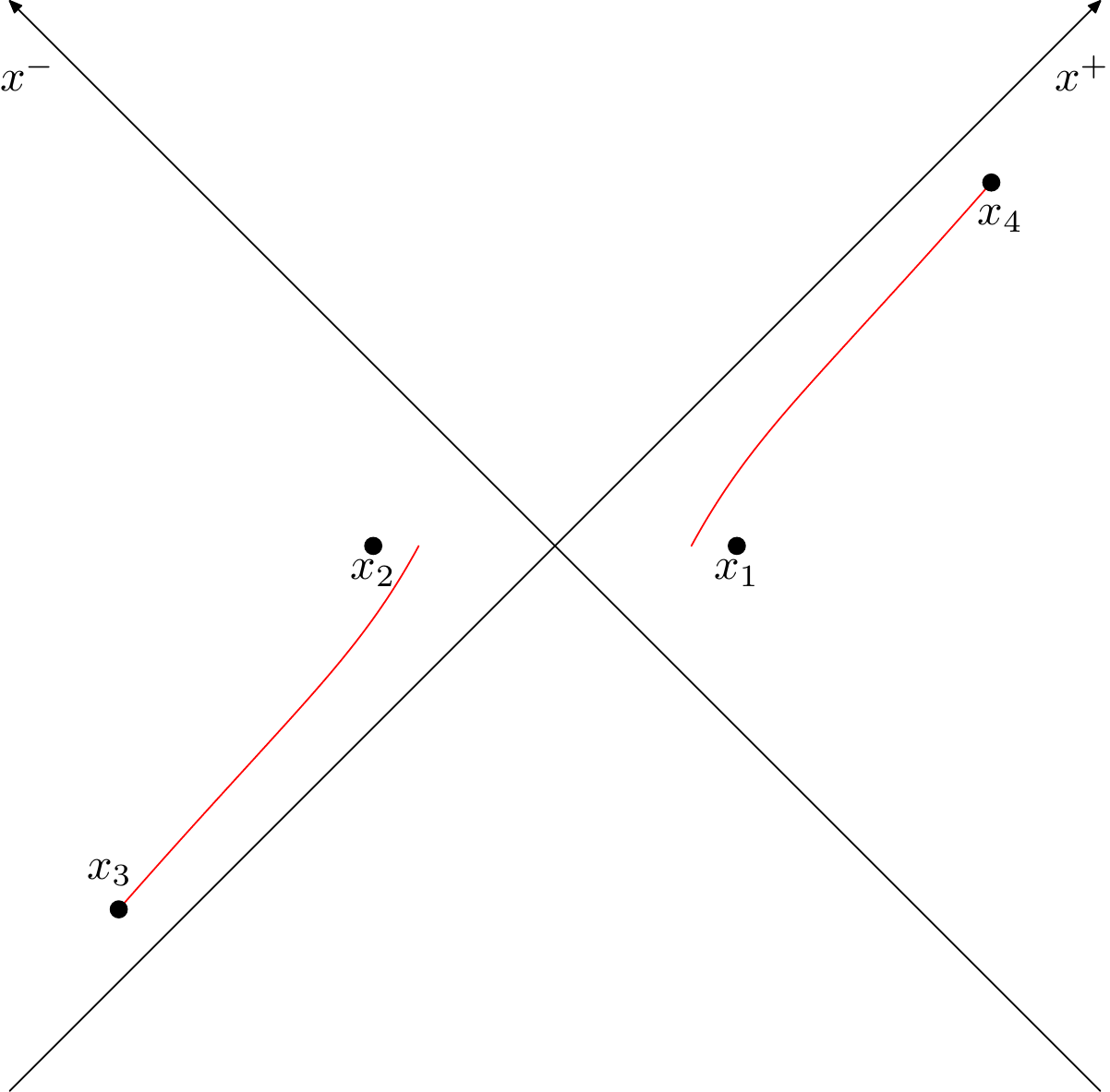}
      \caption{The Regge limit corresponds to taking $t \to \infty$ in (\ref{MSS coordinates}).}
   \label{fig:MSS}   
\end{figure}

We will consider the following kinematics for a four point function, in which we set all four points on the same plane ($x^{\pm} = t \pm x$)
\begin{align} \label{MSS coordinates}
x_1^{\pm} = \pm 1, ~ x_2^{\pm} = \mp 1, ~ x_3^{\pm} = \mp e^{\rho \pm t}, x_4^{\pm} = \pm e^{\rho \pm t}  ,
\end{align}
see figure \ref{fig:MSS}.

The bound on chaos applies for systems at finite temperature with a large number of degrees of freedom. For the case of a large N conformal field theory, a correlation function of single trace primaries $\langle V(x_1) V(x_2) W(x_3) W(x_4)  \rangle$ obeys
\begin{align}\label{statement bound chaos}
\langle V(x_1) V(x_2) W(x_3) W(x_4)  \rangle \approx \langle V(x_1) V(x_2)  \rangle \langle W(x_3) W(x_4)  \rangle (1 + \alpha \frac{e^{\lambda_L t}}{N}),
\end{align}
where the Lyapunov exponent $\lambda_L$ obeys the bound $\lambda_L \leq 2 \pi T$, where $T$ is the temperature of the system. The proportionality constant $ \alpha$ does not depend on $t$. The bound on chaos can be applied to large N CFT's in Minkowski space, in which case we should consider the temperature $T= \frac{1}{2 \pi}$ of the Rindler horizon. 

We cannot apply directly (\ref{statement bound chaos}) to $\langle j_s \jt \jt \jt \rangle$. However, we can use Rindler positivity \cite{Hartman:2016lgu} to bound $\langle j_s \jt \jt \jt \rangle$ by $\langle j_s j_s \jt \jt \rangle$ and $\langle \jt \jt \jt \jt \rangle$ and use the bound on chaos to bound the latter two quantities, as we will explain next.

The Rindler conjugate $\bar{O}$ of an operator $O$ is defined as $\bar{O}_{\mu, \nu ...}(t, x, \vec{y}) = O_{\mu, \nu, ...}^{\dagger} (-t, -x, \vec{y})$, where $\vec{y}$ refers to a transverse coordinate relative to the plane of figure \ref{fig:MSS}. Furthermore we have that $\overline{O_1 O_2} = \bar{O_1} \bar{O_2}$. Rindler positivity and Cauchy-Schwarz inequalities imply that
\begin{align}
|\langle \bar{A} B \rangle|^2 \leq \langle \bar{A} A  \rangle \langle \bar{B} B  \rangle.
\end{align}
where $A$ and $B$ are operators (that might be composite) defined on a single Rindler wedge.

Let us define $A= \jt(x_3) j_s(x_2)$, $B= \jt(x_2) \jt(x_3)$. Then, the time-ordered correlation function in the configuration (\ref{MSS coordinates}) is given by 
\begin{align}
\langle T[j_s(x_1) \jt(x_2) \jt(x_3) \jt(x_4) ] \rangle = \langle \bar{A} B \rangle \\
\leq \sqrt{ \langle \jt(x_4)  j_s(x_1)  \jt(x_3)  j_s(x_2) \rangle \times \langle \jt(x_1)  \jt(x_4)  \jt(x_2)  \jt(x_3) } \rangle \nonumber
\end{align}
The bound on chaos on the rhs of the previous expression implies a bound on $\langle j_s \jt \jt \jt \rangle$. In terms of $\sigma = e^{-t}$:
\begin{align}\label{lambda1}
\lim_{t \rightarrow \infty}\langle T[ j_s(x_1) \jt (x_2) \jt (x_3) \jt (x_4) ] \rangle \sim \frac{\sigma^{ \lambda_1 }}{N} + O(\frac{1}{N^2}),
\end{align}
where $\lambda_1 \geq -1$.

%In order to state the bound, let us introduce another quantity called the Regge intercept (see \cite{Costa:2012cb}), which is related to $\lambda_L$ by $ \lambda_L=j(0)-1$.

% Note that the two point functions $\langle V(x_1) V(x_2)  \rangle$, $\langle W(x_3) W(x_4)  \rangle$ and $\alpha$ do not depend on $t$.

%It is useful to rephrase the bound on $\langle T[j_s(x_1) \jt(x_2) \jt(x_3) \jt(x_4) ] \rangle$ in terms of $\sigma = e^{-t}$: 

%\JS{Is the bound on chaos a theorem? Or is it a conjecture? If it is a conjecture, then our result is also a conjecture :( . This result seems so basic, that it is not clear to me what a proof of it would be.}

\subsection{Consequences for $\langle j_s \jt \jt \jt \rangle_{cb}$}\label{bound on chaos parity even}

%\JS{-------- bad argument: skip -----------------}
%
%Let us define $V= \jt + j_s$ and $W = \jt $. Then, 
%\begin{align}\label{aux bound chaos 1}
%\langle T[j_s(x_1) \jt(x_2) \jt(x_3) \jt(x_4) ] \rangle + \langle T[\jt(x_1) j_s(x_2) \jt(x_3) \jt(x_4) ] \rangle   \\
%= \langle T[V(x_1) V(x_2) W(x_3) W(x_4) ] \rangle - \langle T[j_s(x_1) j_s(x_2) \jt(x_3) \jt(x_4) ] \rangle- \langle T[ \jt(x_1) \jt(x_2) \jt(x_3) \jt(x_4) ] \rangle \nonumber
% \end{align}
%The rhs of (\ref{aux bound chaos 1}) can be bounded using (\ref{statement bound chaos}). Note that for the choice of kinematics (\ref{MSS coordinates}) $\langle j_s(x_1) j_s(x_2)  \rangle$ does not depend on $t$.
%
% Concerning the lhs, both terms must have the same behaviour at large $t$, since the causal configurations are the same in both kinematics \JS{is this a good argument?}. 
% 
% \JS{-------- bad argument: skip -----------------}

%, so the rhs of (\ref{aux bound chaos 1}) contains a dependence like $e^{\lambda_L t}$. 

Let us work out the consequences of the bound on chaos for the Mellin amplitudes of $\langle j_s \jt \jt \jt \rangle$. In the critical boson theory,
\begin{align} \label{Mellin Regge appendix 1}
\langle j_s(x_1) \jt(x_2) \jt(x_3) \jt(x_4) \rangle_{cb}  = |x_1-x_3|^{-4 s-2}  |x_2-x_3|^{2s-1} 
 |x_2-x_4|^{-2s-3}  |x_3-x_4|^{2 s-1}  \\
\times \sum_{k=0}^s \int\int \frac{d \gamma_{12} d{\gamma_{14}}}{(2 \pi i)^2}   \hat{M}(\gamma_{12}, \gamma_{14}; s, k) u^{- \gamma_{12}} v^{-\gamma_{14}} V(1; 2, 3)^{s-k} V(1; 3, 4)^k.  \nonumber
\end{align}
where $V(i; j, k)$ was defined in (\ref{struct V}) and
\begin{align}\label{Mellin Regge appendix 2}
\hat{M}(\gamma_{12}, \gamma_{14}; s, k) = \Gamma(\gamma_{12}) \Gamma(\Delta_1-\gamma_{12}-\gamma_{14}) \Gamma(\gamma_{14}) \Gamma(\gamma_{12} + \frac{\Delta_3 + \Delta_4 - \Delta_1 - \Delta_2}{2}) \\ \Gamma(\frac{\Delta_1 + \Delta_2 - \Delta_3 + \Delta_4}{2}- \gamma_{12} - \gamma_{14}) \Gamma(\gamma_{14} + \frac{\Delta_2 + \Delta_3 - \Delta_1 - \Delta_4}{2}) M(\gamma_{12}, \gamma_{14}; s, k), \nonumber \\
\Delta_1 = 2s+1, ~~~ \Delta_2 = 2, ~~~ \Delta_3 = 2, ~~~ \Delta_4 = 2. \nonumber
\end{align}
We call $M(\gamma_{12}, \gamma_{14}; s, k)$ a Mellin amplitude. The arguments of the $\Gamma$ functions are just the Mellin variables defined in (\ref{def f}).

In the limit $t \rightarrow \infty$ of the kinematics (\ref{MSS coordinates}), the conformal cross-ratio $v$ acquires a monodromy $v \rightarrow v e^{2 \pi i}$. Furthermore
\begin{align}
u \approx 16 \sigma^2 + O(\sigma^3), ~~~ v \approx 1 - 8 \sigma \cosh \rho + O(\sigma^2), ~~ \sigma \rightarrow 0 .
\end{align}

The polynomial growth of the Mellin amplitude is related to the Regge limit, in a manner that we explain next, following appendix C of \cite{Costa:2012cb}. Let us consider the limit
\begin{align}\label{deriving regge Mellin}
\lim_{\sigma \rightarrow 0} \int  \int \frac{d \gamma_{12} d \gamma_{14} }{(2 \pi i)^2}  M(\gamma_{12}, \gamma_{14}; s, k) \Gamma(\gamma_{12}) \Gamma(\Delta_1 - \gamma_{12}- \gamma_{14})  \\
\Gamma(\gamma_{14}) e^{-2 \pi i \gamma_{14}} \Gamma(\gamma_{12} + \frac{\Delta_3 + \Delta_4 - \Delta_1 - \Delta_2}{2})   \Gamma(\gamma_{14} + \frac{\Delta_2 + \Delta_3 - \Delta_1 - \Delta_4}{2}) \nonumber  \\
 \Gamma(-\gamma_{12}-\gamma_{14} + \frac{\Delta_1 + \Delta_2 - \Delta_3 + \Delta_4}{2})  \sigma^{-2 \gamma_{12}} (1- 8 \sigma \cosh \rho)^{-\gamma_{14}}.\nonumber
\end{align}
The factor  $e^{-2 \pi i \gamma_{14}}$ becomes very large in the regime $\gamma_{14} \rightarrow i \infty$. This is cancelled by the exponential decay of the $\Gamma$ functions. Let us suppose that the Mellin amplitude grows polynomially as $\gamma_{14}^{\alpha(s, k)} f(\gamma_{12})$, when $\gamma_{14}$ is large and imaginary and $\gamma_{12}$ is fixed. In this regime we can rewrite (\ref{deriving regge Mellin}) as 
\begin{align}
\sim \int \frac{d \gamma_{12}}{2 \pi i} \Gamma(\gamma_{12}) \Gamma(\gamma_{12} + \frac{\Delta_3 + \Delta_4 - \Delta_1 - \Delta_2}{2}) \sigma^{-2 \gamma_{12}} f(\gamma_{12}) \\
 \int_{M1}^{\infty} \frac{d m_1}{2 \pi} m_1^{-2 -2 \gamma_{12} + \Delta_1 + \Delta_2 + \alpha(s, k)} e^{i m_1 (8 \sigma \cosh \rho + O(\sigma^3))}, \nonumber
\end{align}
where $M_1$ is an irrelevant large number. If we substitute $m_1 \rightarrow \frac{m_1}{\sigma}$ we get that the integral (\ref{deriving regge Mellin}) scales like $\sigma^{1 - \Delta_1 - \Delta_2 - \alpha(s, k)}$. In order to compare (\ref{Mellin Regge appendix 1}) with (\ref{lambda1}), we should furthermore take into account the prefactor and the structures in (\ref{Mellin Regge appendix 1}), which scale with $\sigma$. Our conclusion is that $\alpha(s, k)=1- \lambda_1 - k \leq 2-k$.

We can use the crossing symmetry equations
\begin{align}
\hat{M}(\gamma_{12}, \gamma_{14}; s, k)=
   \sum _{k_2=k}^s (-1)^{k_2} \binom{k_2}{k} \hat{M}(2s + 1-k_2 - \gamma_{12} - \gamma_{14},\gamma_{14}-k+k_2; s, k_2) \label{crossing1 reduced} , \\
\hat{M}(\gamma_{12}, \gamma_{14}; s, k)= \hat{M}(\gamma_{14}, \gamma_{12}; s, s-k)  . \label{crossing2 reduced} 
\end{align}
to derive the following bounds on the polynomial growth of the Mellin amplitude
\begin{empheq}[box=\fbox]{align}
\lim_{\beta \rightarrow \infty} M(\gamma_{12}, \beta \gamma_{14}; s, k) \sim \beta^{\alpha_1(s, k)}, ~~~ \alpha_1(s,k) \leq  2 - k \label{bound on chaos Mellin 1}\\
\lim_{\beta \rightarrow \infty} M(\beta \gamma_{12},  \gamma_{14}; s, k) \sim \beta^{\alpha_2(s, k)}, ~~~ \alpha_2(s,k) \leq  2 - s+k \label{bound on chaos Mellin 2} \\
\lim_{\beta \rightarrow \infty} M(i \beta + \gamma_{12},  - i \beta + \gamma_{14}; s, k) \sim \beta^{\alpha_3(s, k)}, ~~~ \alpha_3(s,k) \leq  2 +s.   \label{bound on chaos Mellin 3}
\end{empheq}
We can apply these bounds to the ansatz (\ref{ansatz Mellin s000}). We conclude that 
\begin{empheq}[box=\fbox]{align}
\lim_{\beta \rightarrow \infty} p(\gamma_{12}, \beta \gamma_{14}; s, k) \sim \beta^{\eta_1(s, k)}, ~~~ \eta_1(s,k) = 2 + 2k + \alpha_1(s, k) \leq  4 + k \label{bound on chaos polynomial 1}\\
\lim_{\beta \rightarrow \infty} p(\beta \gamma_{12},  \gamma_{14}; s, k) \sim \beta^{\eta_2(s, k)}, ~~~ \eta_2(s,k) = 2 + 2s - 2k + \alpha_2(s, k) \leq  4 - k +s \label{bound on chaos polynomial 2} \\
\lim_{\beta \rightarrow \infty} p(i \beta + \gamma_{12},  - i \beta + \gamma_{14}; s, k) \sim \beta^{\eta_3(s, k)}, ~~~ \eta_3(s,k) \leq  4+s.   \label{bound on chaos polynomial 3}
\end{empheq}
 The solution that we found respects this bound.

\subsection{The Regge limit of AdS contact diagrams for the parity even structure in $\langle j_s \jt \jt \jt \rangle$}\label{Regge limit parity even}

\begin{figure}[!t]
\centering
\includegraphics[width=0.45\textwidth]{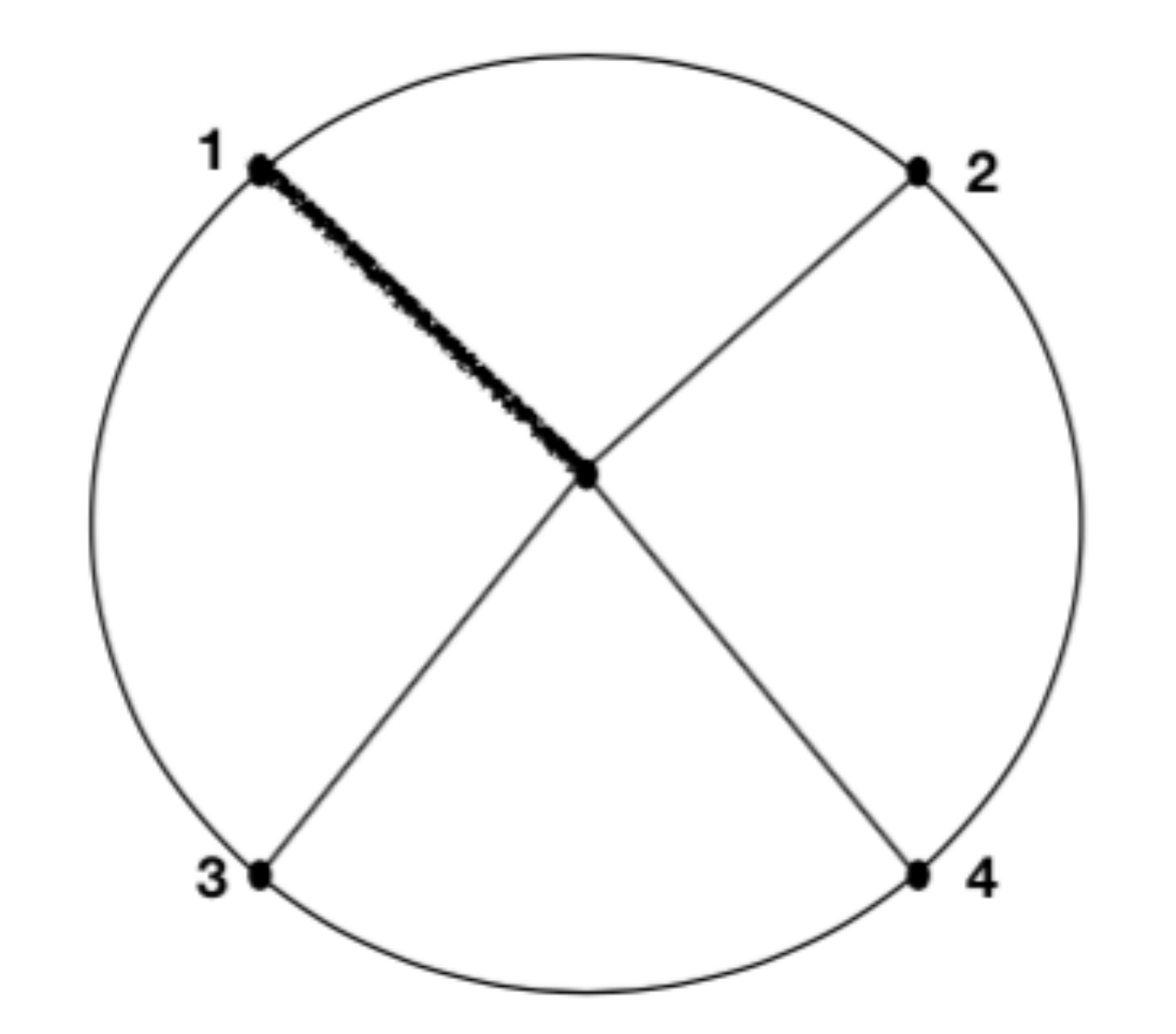}
\caption{AdS contact diagram for $\langle j_s \jt \jt \jt \rangle$.}
\label{fig:contact}
\end{figure}

We will study the Regge limit of a generic AdS contact diagram for $\langle j_s \jt \jt \jt \rangle$ (see figure (\ref{fig:contact})), using the methods of \cite{Costa:2014kfa}. 
We use vectors $P_i$ and $Z_i$ in embedding space to describe the position and polarization vectors of an operator $O_i$ defined on the boundary of AdS. For tensor fields defined on the bulk of AdS, we use vectors $X_i$ and $W_i$ to denote the position and the polarization. The following identities are obeyed:
\begin{align}
Z_i^2 = P_i^2 = Z_i \cdot P_i = X_i^2 +1 = W_i^2= X_i \cdot W_i =0 .
\end{align}

We denote the bulk to boundary propagator of a dimension $\Delta$ and spin $J$ field by $\Pi_{\Delta, J} (X, P; W, Z)$. 	Its formula is
\begin{align}
\Pi_{\Delta, J} (X, P; W, Z) = \mathcal{C}_{\Delta, J} \frac{((-2 P \cdot X)(W \cdot Z) + 2 (W \cdot P) (Z \cdot X)  )^J}{(-2 P \cdot X)^{\Delta+J}  },
\end{align}
where $\mathcal{C}_{\Delta, J}$ is a proportionality constant (whose value will not be relevant for us).

An important class of contact diagrams contributing to the parity even structure in $\langle j_s \jt \jt \jt \rangle$ is given by
\begin{align}\label{integrals AdS}
\int _{AdS} dX ~ \Pi_{\Delta_1=s+1, s_1=s} (X, P_1, K, Z_1) (W \cdot \nabla)^{s_2} \Pi_{\Delta_2=2, s_2=0} (X, P_2) \\  (W \cdot \nabla)^{s_3} \Pi_{\Delta_3=2, s_3=0} (X, P_3) \Pi_{\Delta_4=2, s_4=0} (X, P_4), \nonumber 
\end{align}  
where $s_1 = s_2+s_3$. There are other contact diagrams one can write by contracting more derivatives among the propagators, but such diagrams will diverge more in the Regge limit, which is the issue we wish to discuss here. The covariant derivative is given by
\begin{align}
\nabla_A = \frac{\partial}{\partial X^A} + X_{A} (X \cdot \frac{\partial}{\partial X}) + W_A (X \cdot \frac{\partial}{\partial W} ).
\end{align}
The operator $K$ is given by
\begin{align}
K_A = \frac{d-1}{2}\big( \frac{\partial}{\partial W^A} + X_A (X \cdot \frac{\partial}{\partial W})  \big) + (W \cdot \frac{\partial}{\partial W}) \frac{\partial}{ \partial W^A} \\
+X_A (W \cdot \frac{\partial}{\partial W})(X \cdot \frac{\partial}{\partial W}) - \frac{1}{2} W_A \big( \frac{\partial^2}{\partial W \cdot \partial W} + (X \cdot \frac{\partial}{\partial W})(X \cdot \frac{\partial}{\partial W}) \big), \nonumber
\end{align}
where for our purposes $d=3$.

The following identity
\begin{align}\label{identity propagators}
\Pi_{\Delta_1, s_1} (X, P_1, K, Z_1) (W \cdot \nabla)^{s_2} \Pi_{\Delta_2, s_2} (X, P_2)  (W \cdot \nabla)^{s_3} \Pi_{\Delta_3, s_3} (X, P_3) \\
= C(\Delta_1, \Delta_2, \Delta_3, s_1, s_2, s_3) D_{12}^{s_2} D_{13}^{s_3} \Big( \Pi_{\Delta_1, 0} (X, P_1) \Pi_{\Delta_2 + s_2, 0} (X, P_2) \Pi_{\Delta_3 + s_3, 0} (X, P_3)\Big). \nonumber
\end{align}
is useful for us. $D_{ij}$ is an operator that only acts on the external points. It increases the spin at position $i$ by $1$ and it decreases the conformal dimension at position $j$ by 1. $C(\Delta_1, \Delta_2, \Delta_3, s_1, s_2, s_3)$ is a constant of proportionality, which will not be relevant for us. The precise definition of $D_{ij}$ is
\begin{align}
D_{ij} = (P_j \cdot Z_i ) Z_i \cdot \frac{\partial}{\partial Z_i}  - (P_j \cdot Z_i ) P_i \cdot \frac{\partial}{\partial P_i}  +(P_j \cdot P_i ) Z_i \cdot \frac{\partial}{\partial P_i} .
\end{align}
We confirmed the identity (\ref{identity propagators}) for a few values of the external spins using \textit{Mathematica}. 

So, with the help of identity (\ref{identity propagators}) we can perform the integration in (\ref{integrals AdS}) using only scalar propagators and afterwards we act with the differential operators $D_{12}$ and $D_{13}$. The AdS integral with only scalar propagators corresponds to a contact quartic scalar diagram, whose Mellin amplitude is a constant. Afterwards we act with the differential operators and obtain an expression in the form of (\ref{Mellin s 2}).

Let us exemplify what we mean for the case of $\langle j_2 \jt \jt \jt \rangle$. Let us take $s_2=1$ and $s_3=1$ in (\ref{integrals AdS}). Up to a proportionality constant, the contact diagram is given by
\begin{align}\label{integrals AdS scalar}
D_{12} D_{13} \int _{AdS} dX ~ \Pi_{\Delta_1=3, s_1=0} (X, P_1)  \Pi_{\Delta_2=3, s_2=0} (X, P_2) 
  \Pi_{\Delta_3=3, s_3=0} (X, P_3) \Pi_{\Delta_4=2, s_4=0} (X, P_4)   \\
 \sim D_{12} D_{13} \frac{x_{34}}{x_{23} x_{13}^6  x_{24}^5 } \int\int \frac{d \gamma_{12} d{\gamma_{14}}}{(2 \pi i)^2} \Gamma(\gamma_{12}) \Gamma(3-\gamma_{12}-\gamma_{14}) \Gamma(\gamma_{14})\nonumber \\
  \Gamma(\gamma_{12} - \frac{1}{2}) \Gamma(\frac{5}{2} - \gamma_{12} - \gamma_{14}) \Gamma(\gamma_{14} + \frac{1}{2})    u^{- \gamma_{12}} v^{-\gamma_{14}},     \nonumber 
\end{align}
where the $\sim$ symbol means that we neglected a numerical factor. We now act with the differential operators $D_{12}$ and $D_{13}$ and reorganise the result into the form (\ref{Mellin Regge appendix 1}), (\ref{Mellin Regge appendix 2})\footnote{The step where we gather different terms into the same contour may give rise to subtractions. These do not change our main conclusion, which is that any finite linear combination of AdS contact diagrams for $\langle j_s \jt \jt \jt \rangle$ with $s \geq 4$ does not obey the bound on chaos.}. For this contact diagram, we conclude that
\begin{align}
M(\gamma_{12}, \gamma_{14}, s=2, k=0) = \frac{(-4 + \gamma_{12})(3-8 \gamma_{14} + 4 \gamma_{14}^2)  }{ (-4 + \gamma_{12} + \gamma_{14})  } \\
M(\gamma_{12}, \gamma_{14}, s=2, k=1) = \frac{-2(-2 + \gamma_{12})(-3+ 2 \gamma_{12})(-3+2 \gamma_{14})}{ (-4 + \gamma_{12} + \gamma_{14}) } \nonumber \\
M(\gamma_{12}, \gamma_{14}, s=2, k=2) =\frac{\gamma_{12}(3-8 \gamma_{12} + 4 \gamma_{12}^2)  }{ (-4 + \gamma_{12} + \gamma_{14})  } \nonumber
\end{align}
This contact diagram obeys the chaos bounds (\ref{bound on chaos Mellin 1}), (\ref{bound on chaos Mellin 2}) and (\ref{bound on chaos Mellin 3}). We found that contact diagrams of the type (\ref{integrals AdS}) obey the bound on chaos for spin $2$, but violate the bound on chaos for spin $s \geq 4$.

Our goal is to investigate if there are extra solutions to crossing, conservation and Regge boundedness for $\langle j_s \jt \jt \jt \rangle$. AdS contact diagrams are solutions to the crossing equations, however they are not necessarily conserved, nor Regge bounded. To see that contact diagrams are not necessarily conserved, let us consider a generic contact diagram 
\begin{eqnarray}
\int_{AdS} dX ~ \Pi_{\Delta = s+1, s} (X, P_1, W, Z_1) ~ J(X, P_i, K, Z_i)
\end{eqnarray}
where we denoted by $J(X, P_i, W, Z_i)$ the dependence on the other AdS fields. It turns out that the action of the conservation operator (\ref{conservation operator}) on $\Pi_{\Delta=s+1, s}$ gives a pure gauge expression
\begin{eqnarray}
\frac{\partial}{\partial P} \cdot \mathcal{D}_Z \Pi_{\Delta=s+1, s} (X, P, W, Z) \\
=- 2^{-2-s} s^2 W \cdot \nabla_X \Big( (- P \cdot X )^{-2s-1} \big( (- P \cdot X )(W \cdot Z ) + ( P \cdot W )(X \cdot Z ) \big)^{s-1} \Big)\\
 \equiv  W \cdot \nabla_X F(X, P, W , Z ).
\end{eqnarray}
Thus, 
\begin{eqnarray}
\frac{\partial}{\partial P_1} \cdot \mathcal{D}_{Z_1} \int_{AdS} dX \Pi_{\Delta=s+1, s} (X, P_1, W, Z_1) ~ J(X, P_i, W, Z_i) \\
= - \int_{AdS} dX  F(X, P_1, W , Z_1 )  W \cdot \nabla_X J(X, P_i, K, Z_i)  \nonumber
\end{eqnarray}
This vanishes only if $J(X, P_i, K, Z_i)$ is conserved in the bulk of AdS, i.e. a contact diagram involving a bulk to boundary propagator is conserved only when the bulk to boundary propagator is coupled to a conserved current. Clearly, this is not the case for a generic contact diagram (\ref{integrals AdS}).

So, we consider instead linear combinations of AdS contact diagrams. The most economical way of doing this is to notice that the Mellin transform of any contact diagram, or any linear combination of contact diagrams, can be written as
\begin{align}\label{ansatz Contact Diagrams s000}
\hat{M}(\gamma_{12}, \gamma_{14}; s, k)= \Gamma (-k+\gamma_{14}) \Gamma\left(-k+\gamma_{14}+\frac{1}{2}\right) \Gamma(-s+\gamma_{13}) \\
\times \Gamma\left(-s+\gamma_{13}+\frac{1}{2}\right) \Gamma(k-s+\gamma_{12}) \Gamma\left(k-s+\gamma_{12}+\frac{1}{2}\right) p_{dt}(\gamma_{12}, \gamma_{14}; s, k). \nonumber
\end{align}
where $p_{dt}(\gamma_{12}, \gamma_{14}; s, k)$ is a polynomial. Let us explain this important formula. If we act with the differential operators on the scalar contact diagram, they will shift the arguments of the $\Gamma$ functions by integers. So, the Mellin transform of an AdS contact diagram will involve $6$ $\Gamma$ functions times a polynomial. The arguments of the $\Gamma$ functions are related to the operators that appear in the OPE of the external operators. Thus, we arrive at (\ref{ansatz Contact Diagrams s000}). Notice that $p_{dt}(\gamma_{12}, \gamma_{14}; s, k)$ will eventually have zeros.

The chaos bound for $p_{dt}(\gamma_{12}, \gamma_{14}; s, k)$ is
\begin{align}
\lim_{\beta \rightarrow \infty} p_{dt}(\gamma_{12}, \beta \gamma_{14}; s, k) \sim \beta^{\eta_1(s, k)}, ~~~ \eta_1(s,k) = 2 + 2k + \alpha_1(s, k) \leq  2 + k \label{bound on chaos polynomial double twist 1}\\
\lim_{\beta \rightarrow \infty} p_{dt}(\beta \gamma_{12},  \gamma_{14}; s, k) \sim \beta^{\eta_2(s, k)}, ~~~ \eta_2(s,k) = 2 + 2s - 2k + \alpha_2(s, k) \leq  2 - k +s \label{bound on chaos polynomial double twist 2} \\
\lim_{\beta \rightarrow \infty} p_{dt}(i \beta + \gamma_{12},  - i \beta + \gamma_{14}; s, k) \sim \beta^{\eta_3(s, k)}, ~~~ \eta_3(s,k) \leq  2+s.   \label{bound on chaos polynomial double twist 3}
\end{align}
We imposed crossing and conservation on (\ref{ansatz Contact Diagrams s000}). We find solutions that always violate the chaos bound, for all spins $s \geq 4$. For $s=2$ we find a solution that respects crossing, conservation and Regge boundedness, which is given by
\begin{align}\label{contact diagram spin 2 even}
p_{dt}(\gamma_{12}, \gamma_{14}; s=2, k=0)=\frac{\gamma_{12}^4 \gamma_{14}}{9}+\frac{\gamma_{12}^4}{24}+\frac{\gamma_{12}^3 \gamma_{14}^2}{9}-\frac{5 \gamma_{12}^3 \gamma_{14}}{8}-\frac{5 \gamma_{12}^3}{12}-\frac{7 \gamma_{12}^2 \gamma_{14}^2}{24}\\
+\frac{35 \gamma_{12}^2 \gamma_{14}}{36}+\frac{35 \gamma_{12}^2}{24}+\frac{7 \gamma_{12} \gamma_{14}^2}{72}-\frac{5 \gamma_{12} \gamma_{14}}{24}-\frac{25 \gamma_{12}}{12}+\frac{\gamma_{14}^2}{12}-\frac{\gamma_{14}}{4}+1 \nonumber , \\
p_{dt}(\gamma_{12}, \gamma_{14}; s=2, k=1)= -\frac{2 \gamma_{12}^3 \gamma_{14}^2}{9}+\frac{5 \gamma_{12}^3 \gamma_{14}}{4}-\frac{37 \gamma_{12}^3}{36}-\frac{2 \gamma_{12}^2 \gamma_{14}^3}{9}+\frac{13 \gamma_{12}^2 \gamma_{14}^2}{4}-\frac{331 \gamma_{12}^2 \gamma_{14}}{36} \nonumber \\
+\frac{37 \gamma_{12}^2}{6}+\frac{5 \gamma_{12} \gamma_{14}^3}{4}-\frac{331 \gamma_{12} \gamma_{14}^2}{36}+\frac{77 \gamma_{12} \gamma_{14}}{4}-\frac{407 \gamma_{12}}{36}-\frac{37 \gamma_{14}^3}{36}+\frac{37 \gamma_{14}^2}{6}-\frac{407 \gamma_{14}}{36}+\frac{37}{6} \nonumber , \\
p(\gamma_{12}, \gamma_{14}; s=2, k=2)=\frac{\gamma_{12}^2 \gamma_{14}^3}{9}-\frac{7 \gamma_{12}^2 \gamma_{14}^2}{24}+\frac{7 \gamma_{12}^2 \gamma_{14}}{72}+\frac{\gamma_{12}^2}{12}+\frac{\gamma_{12} \gamma_{14}^4}{9}-\frac{5 \gamma_{12} \gamma_{14}^3}{8} \nonumber \\
+\frac{35 \gamma_{12} \gamma_{14}^2}{36}-\frac{5 \gamma_{12} \gamma_{14}}{24}-\frac{\gamma_{12}}{4}+\frac{\gamma_{14}^4}{24}-\frac{5 \gamma_{14}^3}{12}+\frac{35 \gamma_{14}^2}{24}-\frac{25 \gamma_{14}}{12}+1 . \nonumber
\end{align} 

More explicitly, the spin $2$ parity even contact term in position space is given by 
\begin{align}\label{explicit spin 2 dumb}
\langle j_2 \jt \jt \jt \rangle = |x_1-x_3|^{-10}  |x_2-x_3|^{3} 
 |x_2-x_4|^{-7}  |x_3-x_4|^{3} \sum_{k=0}^s f_k (u, v) w(1; 2, 3)^{s-k} w(1; 3, 4)^k ,
\end{align}
where $w(i; j, k) = \frac{ (x_{ij})_{\mu} x_{ik}^2}{x_{jk}^2} - \frac{(x_{ik})_{\mu} x_{ij}^2}{x_{jk}^2}$ and the indices are symmetrized and their traces removed. For example $w(1; 2, 3)^2 = w(1; 2, 3)_{\mu} w(1; 2, 3)_{\nu} - \frac{\eta_{\mu \nu}}{3}  w(1; 2, 3)^{\rho} w(1; 2, 3)_{\rho}$. Also\footnote{There is some arbitrariness in the choice of the contour. What is important is that it passes to the right of the poles in $\gamma_{12}$, $\gamma_{13}$ and $\gamma_{14}$, see \cite{Penedones:2019tng}.},
\begin{align}
f_k(u, v) = \int_{\frac{7}{3} - k - i \infty}^{\frac{7}{3} - k + i \infty} \frac{d \gamma_{12}}{2 \pi i} \int_{\frac{1}{3} + k - i \infty}^{\frac{1}{3} + k + i \infty} \frac{d\gamma_{14} }{2 \pi i} \hat{M} (\gamma_{12}, \gamma_{14}, s=2, k) u^{-\gamma_{12}} v^{-\gamma_{14}} \label{form f} \\
\hat{M}(\gamma_{12}, \gamma_{14}; s=2, k)= \Gamma (-k+\gamma_{14}) \Gamma\left(-k+\gamma_{14}+\frac{1}{2}\right) \Gamma(3-\gamma_{12} - \gamma_{14}) \label{Mellin spin 2 parity even}  \\
\times \Gamma\left(\frac{7}{2} - \gamma_{12} - \gamma_{14} \right) \Gamma(k-2+\gamma_{12}) \Gamma\left(k+\gamma_{12}-\frac{3}{2}\right) p_{dt}(\gamma_{12}, \gamma_{14}; s=2, k), \nonumber
\end{align} 
where $p_{dt}(\gamma_{12}, \gamma_{14}; s=2, k)$ is given by (\ref{contact diagram spin 2 even}). 

%As to the Mellin contour, we can put the contour in $\gamma_{12}$ from $\frac{7}{3} - k - i \infty$ to $\frac{7}{3} - k + i \infty$ and in $\gamma_{14}$ from $\frac{1}{3} + k - i \infty$ to $\frac{1}{3} + k + i \infty$. 

%The scalar propagators cause a divergence like $ \frac{1}{(z - \bar{z})^{\sum_i \Delta_i -3 + s_1}}$, see formula (\ref{scalar unequal triv}). After acting with the differential operators, we find that the bulk point divergence of the integral (\ref{integrals AdS}) is $ \frac{1}{(z - \bar{z})^{\sum_i \Delta_i -3 + 3 s_1}}$. 

%\JS{we need to consider all the possible linear combinations of the diagrams. afterwards, take the limit $\gamma_{14}\rightarrow \infty$. because it could be that there are cancellations. do this for a few examples, like s1=2, 4, 6, 8, 10}
%
%\JS{Write the contact diagrams in position space. What do they look like?}

\subsection{The Regge limit of AdS contact diagrams for the parity odd structure in $\langle j_s \jt \jt \jt \rangle$}\label{bound on chaos parity odd}

%Our discussion follows closely the one on section (\ref{bound on chaos parity odd}). In this case 

The parity odd structure is
\begin{align} \label{Mellin Regge appendix odd 1}
\langle j_s(x_1) \jt(x_2) \jt(x_3) \jt(x_4) \rangle_{odd}  = |x_1-x_3|^{-4 s-2}  |x_2-x_3|^{2s-2} 
 |x_2-x_4|^{-2s-4}  |x_3-x_4|^{2 s-2}  \\
\times \sum_{k=0}^{s-1} \int\int \frac{d \gamma_{12} d{\gamma_{14}}}{(2 \pi i)^2}   \hat{M}_{odd}(\gamma_{12}, \gamma_{14}; s, k) u^{- \gamma_{12}} v^{-\gamma_{14}} \epsilon(Z_1, P_1, P_2, P_3, P_4) V(1; 2, 3)^{s-1-k} V(1; 3, 4)^k. \nonumber
\end{align}
We define the Mellin amplitude $M_{odd}(\gamma_{12}, \gamma_{14}; s, k)$ in the following manner
\begin{align}\label{Mellin Regge appendix odd}
\hat{M}_{odd}(\gamma_{12}, \gamma_{14}; s, k) = \Gamma(\gamma_{12}) \Gamma(\Delta_1-\gamma_{12}-\gamma_{14}) \Gamma(\gamma_{14}) \Gamma(\gamma_{12} + \frac{\Delta_3 + \Delta_4 - \Delta_1 - \Delta_2}{2}) \\ \Gamma(\frac{\Delta_1 + \Delta_2 - \Delta_3 + \Delta_4}{2}- \gamma_{12} - \gamma_{14}) \Gamma(\gamma_{14} + \frac{\Delta_2 + \Delta_3 - \Delta_1 - \Delta_4}{2}) M_{odd}(\gamma_{12}, \gamma_{14}; s, k), \nonumber \\
\Delta_1 = 2s+1, ~~~ \Delta_2 = 3, ~~~ \Delta_3 = 3, ~~~ \Delta_4 = 3. \nonumber
\end{align}

The following equations encapsulate crossing symmetry:
\begin{align}
\hat{M}_{odd}(\gamma_{12}, \gamma_{14}; s, k)=
   \sum _{k_2=k}^{s-1} (-1)^{k_2} \binom{k_2}{k} \hat{M}_{odd}(2s + 1-k_2 - \gamma_{12} - \gamma_{14},\gamma_{14}-k+k_2; s, k_2) \label{crossing1 odd reduced} , \\
\hat{M}_{odd}(\gamma_{12}, \gamma_{14}; s, k)= \hat{M}_{odd}(\gamma_{14}, \gamma_{12}; s, s-1-k)  . \label{crossing2 odd reduced} 
\end{align}

Let us use the bound on chaos to derive a bound on the polynomial growth of the Mellin amplitude. Let us define the exponent $\alpha(s; k)$ such that $\lim_{\beta \rightarrow \infty} M(\gamma_{12}, \beta \gamma_{14}; s, k) \sim \beta^{\alpha(s; k)}$. In the Regge limit, the Mellin integral goes as $\sigma^{-2s-3-\alpha(s;k)}$. The prefactor times the structure goes as $\sigma^{3+ 2s -k}$. So, (\ref{Mellin Regge appendix odd 1}) behaves as $\sigma^{-k - \alpha(s; k)}$. By comparing with the bound on chaos (\ref{lambda1}) and using (\ref{crossing1 odd reduced}), (\ref{crossing2 odd reduced}) we conclude that 
\begin{align}
\lim_{\beta \rightarrow \infty} M_{odd}(\gamma_{12}, \beta \gamma_{14}; s, k) \sim \beta^{\alpha_1(s, k)}, ~~~ \alpha_1(s,k) \leq  1 - k \label{bound on chaos Mellin 1 odd} \\
\lim_{\beta \rightarrow \infty} M_{odd}(\beta \gamma_{12},  \gamma_{14}; s, k) \sim \beta^{\alpha_2(s, k)}, ~~~ \alpha_2(s,k) \leq  2 - s +k \label{bound on chaos Mellin 2 odd}  \\
\lim_{\beta \rightarrow \infty} M_{odd}(i \beta + \gamma_{12},  - i \beta + \gamma_{14}; s, k) \sim \beta^{\alpha_3(s, k)}, ~~~ \alpha_3(s,k) \leq  s.   \label{bound on chaos Mellin 3 odd}
\end{align}

The Mellin amplitude of an AdS contact diagram of the type (\ref{Mellin Regge appendix odd 1}), or of a linear combination of contact diagrams, is given by
\begin{align}
\hat{M}_{odd}(\gamma_{12}, \gamma_{14}; s, k) = \Gamma(\gamma_{12} +1 + k -s)\Gamma(\gamma_{12} + \frac{1}{2} + k -s) \\ 
\Gamma(\gamma_{14} - k)\Gamma(\gamma_{14} - k - \frac{1}{2})  \Gamma(\gamma_{13} + 1 -s) \Gamma(\gamma_{13} + \frac{1}{2} - s) p_{dt}(\gamma_{12}, \gamma_{14}; s, k), \nonumber
\end{align}
where $\gamma_{13} = 2s + 1 - \gamma_{12} - \gamma_{14}$. The bound on chaos for $p_{dt}(\gamma_{12}, \gamma_{14}; s, k)$ is
\begin{align}
\lim_{\beta \rightarrow \infty} p_{dt}(\gamma_{12}, \beta \gamma_{14}; s, k) \sim \beta^{\lambda_1(s, k)}, ~~~ \lambda_1(s,k) \leq 2+  k \label{bound on chaos polynomial 1 odd} \\
\lim_{\beta \rightarrow \infty} p_{dt}(\beta \gamma_{12}, \gamma_{14}; s, k) \sim \beta^{\lambda_2(s, k)}, ~~~ \lambda_2(s,k) \leq  1 +s - k	 \label{bound on chaos polynomial 2 odd}  \\
\lim_{\beta \rightarrow \infty} p_{dt}(i \beta + \gamma_{12},  - i \beta + \gamma_{14}; s, k) \sim \beta^{\lambda_3(s, k)}, ~~~ \lambda_3(s,k) \leq  1 + s.   \label{bound on chaos polynomial 3 odd}
\end{align}

$p_{dt}(\gamma_{12}, \gamma_{14}; s, k)$ can be found by imposing crossing and conservation. We found that for $s \geq 4$ all solutions violate the bound on chaos. 

However, for $s=2$ there is one solution that respects the bound on chaos. This solution is 
\begin{align}\label{contact diagram spin 2 odd}
p_{dt}(\gamma_{12}, \gamma_{14}; s=2, k=0) = \frac{\gamma_{12}^2}{4}+\frac{\gamma_{12} \gamma_{14}}{2}-\frac{5 \gamma_{12}}{4}-\frac{\gamma_{14}}{2}+1,  \\
p_{dt}(\gamma_{12}, \gamma_{14}; s=2, k=1) = \frac{\gamma_{12} \gamma_{14}}{2}-\frac{\gamma_{12}}{2}+\frac{\gamma_{14}^2}{4}-\frac{5 \gamma_{14}}{4}+1. \nonumber
\end{align}

\section{Open Directions}\label{OpenDirections}

%We believe that the following research directions are worthwhile exploring:
The methods developed in this paper potentially pave the way to compute all four point functions in conformal field theories with slightly broken higher spin symmetry. We believe that the next steps in this program are the following:
\begin{enumerate}
\item \textit{Compute $\langle j_s j_0 j_0 j_0 \rangle$ in the quasi-boson theory.} The conformal structures involved are the same as in this paper, so the calculation should be very similar.

\item \textit{Demonstrate that AdS contact diagrams are not present in $\langle \jt \jt \jt \jt \rangle$ and $\langle j_2 \jt \jt \jt \rangle$ in the quasi-fermion theory using pure CFT arguments.} The chaos bound allows for contact diagrams in $\langle \jt \jt \jt \jt \rangle$ and $\langle j_2 \jt \jt \jt \rangle$. Their absence for $\langle \jt \jt \jt \jt \rangle$ was demonstrated in \cite{Turiaci:2018dht} using Feynman diagrams. It should be possible to give a pure CFT demonstration of this fact. The idea is to write down the higher spin Ward identity that connects $\langle \jt \jt \jt \jt \rangle$ and $\langle j_2 \jt \jt \jt \rangle$, plug the AdS contact diagrams multiplied by arbitrary functions of the t'Hooft coupling and obtain that the only way for the Ward identity to be satisfied is if such functions vanish.

\end{enumerate}
Let us mention some more ambitious problems:
\begin{enumerate}
\item \textit{Develop a code that computes all spinning four point functions in CFT's with slightly broken higher spin symmetry.} Such a code should:
\begin{itemize}

\item  generate the structures involved for a given four point function

\item  generate an ansatz for the Mellin transform, which should be a product of $6$ Gamma functions (whose arguments are determined by the lightcone OPE, which is known) times polynomials

\item impose crossing, pseudo-conservation and Regge boundedness to fix all the undetermined coefficients in the polynomials.

\end{itemize}

What differs from what we did here is that for generic spins we should \underline{not} use embedding space, since the conformal structures in embedding space are generically linearly dependent on each other. It is best to use conformal frame techniques instead. Concretely, one would need the $3$ dimensional version of \cite{Cuomo:2017wme} (see also \cite{Kravchuk:2016qvl}).

\item \textit{Demonstrate that AdS contact diagrams are not present in four point functions in CFT's with slightly broken higher spin symmetry}. As above, the hurdle should be in adapting our formalism to use the 3d conformal frame.

% \item \textbf{Understand CFT's with slightly broken higher spin symmetry from the point of view of the bulk of AdS.} This is the most ambitious problem, and the ultimate reason to study such CFT's in the first place. Recently, in \cite{Aharony:2020omh} the path integral for the critical boson theory was rewritten in terms of higher spin fields in AdS, thus proving AdS/CFT for this theory. It would be good to understand how \cite{Aharony:2020omh} connects with our approach. A concrete goal would be to think how to generalise the methods of \cite{Aharony:2020omh} to generic CFT's with slightly broken higher spin symmetry at finite t'Hooft coupling.
% 
%  \item \textbf{Understand the breaking of higher spin symmetry at the classical level.} Let us suppose that one has fully demonstrated the AdS/CFT correspondence for the case of CFT's with slightly broken higher spin symmetry. This serves as a warmup problem to study the duality between $\mathcal{N}=4$ SYM and type IIB string theory. In this case the higher spin symmetry is broken at the classical level, i. e. one finds single trace operators on the rhs of the pseudo-conservation equations. It should take a long time before one gets to this ultimate goal.

\end{enumerate}

Recently, a new formalism for correlators of conserved currents was proposed in \cite{Caron-Huot:2021kjy}. The idea is to write the conformal structures in a helicity basis. It would be very interesting to apply this idea to correlators in CFT's with slightly broken higher spin symmetry. 

Ultimately, one would like to understand higher spin symmetry from the point of view of the bulk of AdS. We hope that our CFT computations can be of some utility for this ultimate goal.

\textit{Acknowledgments}. I am very grateful to Joao Penedones and Alexander Zhiboedov for suggesting me this problem and for all the help they provided me. All the errors are of course mine. Furthermore, I would like to thank discussions with Aditya Hebbar, Evgeny Skvortsov, Subham Chowdhury and Vasco Gonçalves. This work  was partially supported by a grant from the Simons Foundation (Simons Collaboration on the Nonperturbative Bootstrap: 488649) and by the Swiss National Science Foundation through the project 200021-169132. This project has received funding from the European Research Council (ERC) under the European Union's Horizon 2020 research and innovation programme (grant agreement number 949077).

\appendix

\section{Bulk Point Limit}\label{app: bulk}

Correlation functions of conformal field theories in Lorentzian signature may diverge even when none of the distances between the points vanish. At the moment a full classification of the singularity structure of correlation functions in conformal field theories does not exist.

%One such singularity is the so called ``bulk point singularity". It appears in the following manner. Consider a four point function in a conformal field theory and imagine that such a conformal field theory lives in the boundary of AdS. Sometimes the configuration of the external points is such that there is a point in the interior of AdS that is null separated from all four external points. In that case, the correlation function diverges. Such a singularity is analogous to a Landau singularity for scattering amplitudes. A detailed examination of the bulk point limit for a four point function of equal scalars was carried out in \cite{Maldacena:2015iua}.

%We emphasize that the CFT does not need to have an AdS dual in order to exhibit a bulk point singularity.

One such singularity is the so called ``bulk point singularity". In terms of cross ratios, we can obtain such a singularity in the following manner. In Lorentzian signature $z$ and $\bar{z}$ are independent real numbers. The four point function has branch points. When $z$ and $\bar{z}$ go around the branch points the four point function may develop a divergence when $z = \bar{z}$. More specifically, suppose $z$ goes around the branch point at $1$, $\bar{z}$ goes around $\infty$ and now take $z \rightarrow \bar{z}$. We generically expect the four point function to diverge in this limit. A detailed examination of the bulk point limit for a four point function of equal scalars was carried out in \cite{Maldacena:2015iua}.

%say what is the point of this section. to state our expectation for the bulk point limit of <s000>. this is an important ingredient in our discussion.

In the bulk point limit a $d$ dimensional conformal block where the external operators are scalars diverges as $\frac{1}{(z - \bar{z})^{d-3}}$ \cite{Maldacena:2015iua}. For this reason it is expected that a generic nonperturbative four point function of scalars diverges as
\begin{align}
\langle \O \O \O \O \rangle  \sim \frac{1}{(z - \bar{z})^{d-3}}.
\end{align}

However, when the CFT has a local bulk dual, then we expect the divergence to be more severe. For example, a contact quartic diagram in AdS diverges as
\begin{align} \label{bulk point contact scalar}
\langle \O \O \O \O \rangle  \sim \frac{1}{(z - \bar{z})^{4 \Delta -3}}.
\end{align}

The plan for this section is the following. In \ref{app: bulk unequal} we calculate the bulk point singularity of an AdS contact diagram for a scalar four point function of unequal primaries. The result is a trivial generalisation of (\ref{bulk point contact scalar}), however to our knowledge its derivation had not appeared before in the literature. We need such a result in order to calculate the bulk point singularity of an AdS contact diagram for $\langle j_s \jt \jt \jt \rangle$, which we do in section \ref{app: bulk contact diagrams in AdS}. Finally, in section \ref{app: bulk in HSS} we calculate the expected bulk point divergence of $\langle j_s \jt \jt \jt \rangle$ in CFT's with slightly broken higher spin symmetry. We assume that $\langle j_s \jt \jt \jt \rangle$ does not diverge more than conformal blocks in the bulk point limit. We conclude that AdS contact diagrams diverge more severely in the bulk point limit than what is expected for $\langle j_s \jt \jt \jt \rangle$ for $s \geq 2$ in CFT's with slightly broken higher spin symmetry. Thus, bulk point softness implies that we cannot add AdS contact diagrams to the solution to the pseudo-conservation equations that we found in section (\ref{algorithm in Mellin space}).

Let us add a caveat. Our result for $\langle j_s \jt \jt \jt \rangle$ does not rely on assuming bulk point softness and is independent of it. Nevertheless, we choose to keep this appendix, because it was useful for us to think in terms of the bulk point limit in the early stages of our work, and maybe this can be of use to someone else.

%$\langle j_s \jt \jt \jt \rangle$ in CFT's with slightly broken higher spin symmetry, which we do in section \ref{app: bulk in HSS}. Finally in section \ref{app: bulk contact diagrams in AdS} we compute the bulk point divergence of an AdS contact diagram for $\langle j_s \jt \jt \jt \rangle$. We show that such a divergence is more severe than what we expect for $\langle j_s \jt \jt \jt \rangle$ in CFT's with slightly broken higher spin symmetry.

\subsection{Bulk point singularity of an AdS contact diagram for a scalar four point function of unequal primaries} \label{app: bulk unequal}

A quartic contact diagram has a Mellin amplitude equal to $1$. We will use this to compute the bulk point divergence, proceeding similarly to section $7.5.1$ in \cite{Penedones:2019tng}. Upon analytic continuation, the diagram is given by
\begin{align}
\frac{ \langle \O_1  \O_2  \O_3  \O_4  \rangle }{p} = \int \int \frac{d \gamma_{12} d \gamma_{14}}{(2 \pi i)^2} \Gamma(\gamma_{12}) \Gamma(\gamma_{13}) \Gamma(\gamma_{14}) \Gamma(\gamma_{12} + a_{34}) \Gamma(\gamma_{13} + a_{24}) \Gamma(\gamma_{14} + a_{23}) u^{- \gamma_{12}} v^{- \gamma_{14}} \\
\rightarrow \int \int \frac{d \gamma_{12} d \gamma_{14}}{(2 \pi i)^2} \Gamma(\gamma_{12}) \Gamma(\gamma_{13}) \Gamma(\gamma_{14}) \Gamma(\gamma_{12} + a_{34}) \Gamma(\gamma_{13} + a_{24}) \Gamma(\gamma_{14} + a_{23}) u^{- \gamma_{12}} v^{- \gamma_{14}} e^{-2 \pi i (\gamma_{12} + \gamma_{14})} \nonumber, \\
p= |x_1-x_3|^{-2 \Delta_1} |x_2 - x_3|^{-2 a_{23}} |x_2 - x_4|^{-2 a_{24}-2\Delta_1} |x_3 - x_4|^{-2 a_{34}} \nonumber
\end{align}
where $a_{ij} =2( \Delta_i + \Delta_j )- \sum_k  \Delta_k$ and $\gamma_{13}= \Delta_1 - \gamma_{12} - \gamma_{14}$. The integral diverges when $\gamma_{12}$ and $\gamma_{14}$ have a very big and positive imaginary part. We can use Stirling's approximation for the $\Gamma$ functions. Indeed suppose we take $\gamma_{12} = i s \beta$ and $\gamma_{14} =i  s (1-\beta)$. Then for very large $s$ we have
 \begin{align}
\langle \O_1  \O_2  \O_3  \O_4 \rangle  \approx p \int_{s_0}^{\infty} \frac{ds}{s} \int_{0}^1 d \beta  s^{\frac{\sum_i \Delta_i}{2} -1 } f(\beta) \\
\times \exp \Big( i s  \big( -2  (\beta -1) \log (1-\beta )+2  \beta  \log (\beta )- \beta  \log (u)+(\beta -1) \log (v)  \big) \Big),   \nonumber
\end{align}
where $f(\beta)$ is a function of $\beta$ that will not play any role. The integral has a saddle point for $\beta \to \beta_s = \frac{\sqrt{u}}{\sqrt{u}+\sqrt{v}}$. In that case the exponential dependence of the integrand becomes $e^ {i s \left(\frac{\left(\sqrt{u}+\sqrt{v}\right)^2}{\sqrt{u} \sqrt{v}} (\beta-\beta_s)^2  -2 \log \left(\sqrt{u}+\sqrt{v}\right)\right)}$. The integral in $\beta$ is Gaussian and can be readily evaluated. Furthermore, the phase is stationary when $\sqrt{u} + \sqrt{v} =1$. In that case we have $\log( \sqrt{u} + \sqrt{v} ) \sim (z - \bar{z})^2$. So, we conclude that  
\begin{align}\label{scalar unequal triv}
\langle \O_1  \O_2  \O_3  \O_4  \rangle \sim \int_{s_0}^{\infty} \frac{ds}{s} s^{\frac{\sum_i \Delta_i}{2} - \frac{3}{2} } e^{i s (z - \bar{z})^2} \sim \frac{1}{(z - \bar{z})^{\sum_i \Delta_i - 3}}. 
\end{align}

\subsection{Bulk point singularity of AdS contact diagrams for $\langle j_s \jt \jt \jt \rangle$}\label{app: bulk contact diagrams in AdS}

Identity (\ref{identity propagators}) allows us to obtain spinning contact AdS diagrams from scalar contact AdS diagrams. So, with the help of identity (\ref{identity propagators}) we can perform the integration in (\ref{integrals AdS}) using only scalar propagators and afterwards we act with the differential operators $D_{12}$ and $D_{13}$. The scalar propagators cause a divergence like $ \frac{1}{(z - \bar{z})^{\sum_i \Delta_i -3 + s}}$, see formula (\ref{scalar unequal triv}). After acting with the differential operators, we find that the bulk point divergence of the integral (\ref{integrals AdS}) is $ \frac{1}{(z - \bar{z})^{\sum_i \Delta_i -3 + 3 s}} = \frac{1}{(z- \bar{z})^{4 s +4}}$. 

\subsection{Bulk point singularity of $\langle j_s \jt \jt \jt \rangle$ in CFT's with slightly broken higher spin symmetry}\label{app: bulk in HSS}

Conformal field theories with slightly broken higher spin symmetry have an infinite number of light single trace operators. For this reason, they are not expected to be dual to a local theory in AdS. Thus, their bulk point singularity should not be enhanced with respect to that of an individual conformal block.

We want to calculate the bulk point divergence of $\langle j_s  \jt \jt \jt \rangle$. For our discussion, it is useful to introduce the operator 
\begin{align}
d_{11} = ( P_1 \cdot P_2 ) Z_1 \cdot \frac{\partial}{\partial P_2}  - ( Z_1 \cdot P_2 ) P_1 \cdot \frac{\partial}{\partial P_2}  - ( Z_1 \cdot Z_2 ) P_1 \cdot \frac{\partial}{\partial Z_2}  + ( P_1 \cdot Z_2 ) Z_1 \cdot \frac{\partial}{\partial Z_2},
\end{align}
where we used embedding space coordinates \cite{Costa:2011mg}. This operator acts on conformal blocks where the operator exchanged is symmetric and traceless. It increases the spin of the operator in position $1$ by $1$ and it decreases its conformal dimension by $1$ also. It turns out that $d_{11}^s ( z - \bar{z})^a  \sim ( z - \bar{z})^{a-2s}$, i.e. the action of $d_{11}^s$ increases the divergence by a power of $2s$. For this reason, we expect the divergence of $\langle j_s  \jt \jt \jt \rangle$ to be
\begin{align}\label{bulk point HSS appendix}
\langle j_s  \jt \jt \jt \rangle \sim \frac{1}{(z - \bar{z})^{2 s}}
\end{align}
since the scalar conformal block diverges logarithmically. We could have picked other differential operators than $d_{11}$ to create spin from the scalar conformal block. Since such operators only contain first derivatives of $P_i$ (and not higher derivatives), they lead to the same divergence (\ref{bulk point HSS appendix}).

\section{Algorithm for computing $\langle j_s \jt \jt \jt \rangle$ in position space}\label{app:algorithm in position space}

We will implement an algorithm in position space to calculate $\langle j_s \jt \jt \jt \rangle_{cb}$. The results match with the Mellin space calculation. 

%This is important, since the position space calculation is sensitive to terms that have a vanishing Mellin transform, like powers of $u$ or powers of $v$. The Mellin space calculation is not sensitive to such terms.	

%\textbf{Ansatz}. $\langle j_s \jt \jt \jt \rangle_{cb}$ is constrained by conformal symmetry, crossing, consistency with OPE and the pseudo-conservation equation that $j_s$ obeys. Conformal symmetry implies that
%\begin{align}
%\langle j_s \jt \jt \jt \rangle = p \sum_{j=0}^s f_j(u, v) w(1; 2, 3)^j w(1; 3, 4)^{s-j},
%\end{align}
$\langle j_s \jt \jt \jt \rangle_{cb}$ is constrained by conformal symmetry, crossing, consistency with OPE and the pseudo-conservation equation that $j_s$ obeys. Conformal symmetry implies that
\begin{align}
\langle j_s \jt \jt \jt \rangle_{cb} = p \sum_{j=0}^s f_j(u, v) w(1; 2, 3)^j w(1; 3, 4)^{s-j},
\end{align}
where 
\begin{align}
p \equiv \frac{ (x_{23}^2 x_{24}^2 x_{34}^2)^{\frac{s}{3} - \frac{5}{6}} }{(x_{12}^2 x_{13}^2 x_{14}^2)^{\frac{2s}{3} + \frac{1}{3}}  }, ~ u \equiv \frac{x_{12}^2 x_{34}^2}{x_{13}^2 x_{24}^2}, ~ v \equiv \frac{x_{14}^2 x_{23}^2}{x_{13}^2 x_{24}^2}, \\
w(i; j, k) \equiv (x_{ij})_{\mu} \frac{x_{ik}^2}{x_{jk}^2}-(x_{ik})_{\mu} \frac{ x_{ij}^2}{x_{jk}^2} \nonumber
\end{align}
and we use the notation $(x_{ij})_{\mu}= (x_i)_{\mu}- (x_j)_{\mu}$, $x_{ij}=|x_i - x_j|$. The indices are symmetric and traceless. $f_j(u,v)$ is a function of the cross ratios not determined by conformal symmetry. 

We write the following ansatz. 
\begin{align}\label{ansatz}
f_j(u, v) = \frac{u^{a(j)} v^{b(j)}}{(1+\sqrt{u} + \sqrt{v})^{s}} \sum_{n_j=0}^{N(j)} \sum_{m_j=0}^{M(j)} c_{n_j, m_j} u^{\frac{n_j}{2}} v^{\frac{m_j}{2}},
\end{align}
where
%\begin{align}
%a(j)= bla , ...\label{consistency with OPE}
%\end{align}
$c_{n_j, m_j}$ are parameters that will be fixed by crossing and the pseudo-conservation equation. The values of $a(j)$, $b(j)$, $M(j)$ and $N(j)$ will follow from consistency with the operator product expansion.  

Let us motivate the preceding ansatz. The spinning four point functions are related to the scalar four point functions by slightly broken higher spin Ward identities. The scalar four point function is a linear combination of powers of $u$ and of $v$. So, it is natural that $f_j(u, v)$ is made up of powers of $u$ and of $v$. 

We will see below that the contribution to the operator product expansion of a certain operator goes as $\sim u^{\frac{\tau}{2}}$, where $\tau$ is the twist, which is defined as the conformal dimension minus the spin. Since all operator dimensions are integers, it is natural that the ansatz involves semi-integer powers of $u$ and of $v$. The denominator $\frac{1}{(1 + \sqrt{u}+ \sqrt{v})^{s}}$ diverges in the bulk point limit as $\frac{1}{(z-\bar{z})^{2s}}$, which agrees with the discussion in \ref{app: bulk in HSS}.

%The most nontrivial part is the denominator $\frac{1}{(1 + \sqrt{u}+ \sqrt{v})^{s}}$ in (\ref{ansatz}). From \cite{Maldacena:2015iua} we expect correlators to have a Landau singularity. The denominator $\frac{1}{(1 + \sqrt{u}+ \sqrt{v})^{s-1}}$ does precisely that. \cite{Maldacena:2015iua} only analyses scalar correlators, so we did not know the precise power $s-1$ beforehand, it came out of our calculation. The case $s=2$ is an exception, in that case the denominator is absent. 
%The expression $1 + \sqrt{u}+ \sqrt{v}$ also appears in $\langle \sigma \sigma \sigma \sigma \rangle$ in the 2d Ising model \cite{Alday:2015ota}, so it was natural to try it here.

%$O_{\mu_1, ... \mu_{l_1}}(x)$, $O_{\nu_1, ... \mu_{l_2}}(0)$

We can fix $a(j), b(j), N(j), M(j)$ by consistency with the lightcone operator product expansion. Let us explain the general idea. Consider two primary operators $O_{\mu_1 ... \mu_{l_1}}(x)$, $O_{\nu_1 ... \nu_{l_2}}(0)$ of conformal dimensions $\Delta_1$ and $\Delta_2$ and spins $l_1$ and $l_2$ and suppose they exchange a primary operator $O_{\rho_1... \rho_{l}}$ of dimension $\Delta$ and spin $l$. The most singular term  due to $O_{\rho_1... \rho_{l}}$ that can appear in the lightcone operator product expansion is $\frac{O_{\rho_1 ... \rho_{l}} x^{\rho_1} ... x^{\rho_l}x_{\{ \mu_1} ... x_{\mu_{l_1} \} } x_{\{ \nu_1} ... x_{\nu_{l_2} \} }  }{|x|^{\Delta_1 + \Delta_2 + l_1 + l_2  + l - \Delta}} \sim (x^2)^{- \frac{\Delta_1 + \Delta_2 + l_1 + l_2}{2} + \frac{\tau}{2}} $, where the $\mu$ and $\nu$ indices are traceless symmetric and $\tau= \Delta-l$. 

For $\langle j_s \jt \jt \jt \rangle$ the primary operators exchanged can have twist $1$ (higher spin currents), $3 + 2n$ (double traces $[j_s, \jt]$) and $4 + 2n$ (double traces $[\jt, \jt]$), where $n$ is a nonnegative integer. There is no primary operator of twist $2$ being exchanged. This is an important condition that we impose in our algorithm. 

More explicitly 
\begin{align} \label{OPE bounds}
j_s(x) \jt(0) \sim (x^2)^{-s-1} j_{s'} + 
(x^2)^{-s} [j_s, \jt] + (x^2)^{-s + \frac{1}{2}} [\jt, \jt],  \\ 
\jt(x) \jt(0) \sim (x^2)^{- \frac{3}{2}} j_{s'} \label{OPE bounds 2} 
+ (x^2)^{- \frac{1}{2}} [j_s, \jt] + (x^2)^{0} [\jt, \jt] ,
\end{align}
where we wrote the most singular powers of the distance that can appear in the lightcone operator product expansion. Our ansatz (\ref{ansatz}) needs to be compatible with (\ref{OPE bounds}), (\ref{OPE bounds 2}). This fixes $a(j), b(j), N(j), M(j)$.

The final ingredient is compatibility with pseudo-conservation. $\partial \cdot j_s$ can have contributions coming from $[j_{s_1}, \jt ]$ and $[j_{s_1}, j_{s_2} ]$. Only the former matter since we are interested in $\langle j_s \jt \jt \jt \rangle$. More precisely,
\begin{align}
\partial \cdot j_s \supset \sum_{s_1=2}^{s-2} \sum_{m=0}^{s-s_1-1} c_m \partial^m j_{s_1} \partial^{s-s_1-1-m} \jt.
\end{align}
Since the right-hand side must be a conformal primary, this implies \cite{Giombi:2017rhm} 
\begin{align}\label{primary check}
c_m = \frac{- (m-s+s_1)(m-s+s_1-1)}{m(m+2 s_1)} c_{m-1}.
\end{align} 
Thus $\langle \partial \cdot j_s \jt \jt \jt \rangle$ is a linear combination of terms of type $\partial^{n_1} \langle \jt \jt \rangle \partial^{n_2} \langle  j_{s_1} \jt \jt \rangle $.

Crossing and compatibility with pseudo-conservation fix all coefficients in (\ref{ansatz}) up to a number. This number is related to the normalizaton of $j_s$. In fact we did not even need to input formula (\ref{primary check}), we kept the coefficients $c_m$ as unknowns and our algorithm correctly returns (\ref{primary check}). This serves as a check on our results. We checked that the algorithm fixes the solution for $s=2, ..., 14$. Afterwards the computation becomes heavy for our laptop.

\section{Mixed Fourier Transform}\label{app: mixed Fourier}

We will solve the higher spin Ward identities to compute $\langle j_2 \jt \jt \jt \rangle$. This is a rederivation of the main result of \cite{Li:2019twz}. Our method involves the use of a mixed Fourier transform, see \cite{Caron-Huot:2021kjy} and \cite{Jain:2020puw}.

We use the metric $ds^2 = - dx^- dx^+ + dy^2$. We will take all indices lowered and in the minus component. We will study the action of the charge
\begin{align}
Q = \sqrt{\tilde{N}} \alpha_4 \int_{x^{+}= const.} dx^- dy j_{----}
\end{align}
on the four point function $\langle \jt \jt \jt \jt \rangle$. We make use of equations \cite{Maldacena:2012sf}, \cite{Turiaci:2018dht}
\begin{align}
\partial \cdot j_4 = \alpha \frac{\tilde{\lambda}}{\sqrt{\tilde{N}} \sqrt{1+\tilde{\lambda}^2}} (:\partial_{-} j_{\tilde{0}} j_2: - \frac{2}{5} : j_{\tilde{0}} \partial_{-} j_2 :), \label{divj4} \\
\lbrack Q, j_{\tilde{0}} \rbrack =  \partial_{-}^3 j_{\tilde{0}} + \frac{\beta}{\sqrt{1+\tilde{\lambda}^2}} (\partial_{-}\partial_{-} j_{- y} - \partial_{-}\partial_{y} j_{--}). \label{Q comm}
\end{align}
$\alpha$, $\alpha_4$ and $\beta$ are numerical coefficients that can be obtained from solving Ward identities at the level of three point functions\footnote{We normalised the charge such that the coefficient multiplying $\partial_{-}^3 j_{\tilde{0}}$ in (\ref{Q comm}) is $1$.}. We will not need their precise value in what follows.

The scalar four point function obeys the slightly broken spin $4$ Ward identity
\begin{align}\label{explicit}
\langle [Q, \jt] \jt \jt \jt \rangle + ... 
= \sqrt{\tilde{N}} \alpha_4 \int d^3 x \langle \partial \cdot j_4(x) \jt \jt \jt \jt \rangle, 
\end{align}
where by $...$ we mean the permutations $(12)$, $(13)$, $(14)$. Note that
\begin{eqnarray}
\langle \jt \jt \jt \jt \rangle= \langle \jt \jt \jt \jt \rangle_{disc} + \frac{1}{N}\langle \jt \jt \jt \jt \rangle_{ff},
\end{eqnarray}
 where $\langle \jt \jt \jt \jt \rangle_{ff}$ denotes the connected piece in the free fermion theory and $\langle \jt \jt \jt \jt \rangle_{disc}$ denotes the disconnected piece. The disconnected piece obeys
\begin{eqnarray}
\langle \partial^3 \jt \jt \jt \jt \rangle_{disc} + ... = 0,
\end{eqnarray}
where we summed over all permutations. For this reason the disconnected piece drops out of (\ref{explicit}). Using our ansatz (\ref{n and l}) we conclude that
\begin{align} 
\langle [Q, \jt] \jt \jt \jt \rangle + ... =
\frac{1}{\tilde{N}}\langle \partial^3 \jt \jt \jt \jt \rangle_{ff} 
+ \frac{\beta}{\tilde{N} (1 + \tilde{\lambda}^2)} \big(\langle (\partial_{-}\partial_{-} j_{- y} - \partial_{-}\partial_{y} j_{--}) \jt \jt \jt \rangle_{ff} \\
 + \tilde{\lambda} \langle (\partial_{-}\partial_{-} j_{- y} - \partial_{-}\partial_{y} j_{--}) \jt \jt \jt \rangle_{cb} \big) + ...\nonumber
\end{align}
From the Ward identities in the free fermion theory this becomes
%\begin{eqnarray}
%\langle \partial^3 \jt \jt \jt \jt \rangle_{ff} + \beta \langle (\partial_{-}\partial_{-} j_{- y} - \partial_{-}\partial_{y} j_{--}) \jt \jt \jt \rangle_{ff}+ perm. =0,
%\end{eqnarray}
%up to contact terms. We found that $\beta=\frac{18}{5}$ \footnote{When we were using the Euclidean metric, we had found that $\beta= \frac{18 i}{5}$. Our whole notebook "WI using vector calculus" has this issue.}.
%Now suppose we use the Lorentzian metric (see Sasha's notes for conventions). Then, we see that in $+, -$ and $y$ coordinates, the epsilon tensor has no factor of $i$. Indeed, it is $-i$ times the previous object. Then, $\beta_{Lorentzian}=i \beta_{Euclidean}= - \frac{18}{5}$
%Thus, working up to contact terms, our final expression for the lhs of the slightly broken Ward identities is
\begin{align} \label{mix fourier intermediate 0}
\langle [Q, \jt] \jt \jt \jt \rangle + ... = - \frac{\tilde{\lambda}^2 \beta}{\tilde{N}(1+\tilde{\lambda}^2)} \langle (\partial_{-}\partial_{-} j_{- y} - \partial_{-}\partial_{y} j_{--}) \jt \jt \jt \rangle_{ff} \\
 + \frac{\tilde{\lambda} \beta}{\tilde{N}(1+\tilde{\lambda}^2)}  \langle (\partial_{-}\partial_{-} j_{- y} - \partial_{-}\partial_{y} j_{--}) \jt \jt \jt \rangle_{cb} + ... \nonumber 
\end{align}

Using (\ref{divj4}) in the right-hand side of (\ref{explicit}) we get
\begin{align}\label{mix fourier intermediate 1}
\sqrt{\tilde{N}} \alpha_4 \int d^3 x \langle \partial \cdot j_4(x) \jt \jt \jt \jt \rangle = \alpha \alpha_4 \frac{\tilde{\lambda}}{\sqrt{1+ \tilde{\lambda}^2}} \int d^3 x \big(\langle \partial_- \jt (x) \jt \rangle \langle j_2 (x) \jt \jt \jt \rangle \\
- \frac{2}{5} \langle \jt(x) \jt \rangle \langle \partial_- j_2(x) \jt \jt \jt \rangle + ...  \big)\nonumber
\end{align}
We use the decomposition (\ref{n and l}) to obtain that (\ref{mix fourier intermediate 1}) is equal to
\begin{align}
\alpha \alpha_4 \frac{\tilde{\lambda}}{\tilde{N} (1+ \tilde{\lambda}^2)} \int d^3 x \big(\langle \partial_- \jt (x) \jt \rangle \langle j_2 (x) \jt \jt \jt \rangle_{ff}
- \frac{2}{5} \langle \jt(x) \jt \rangle \langle \partial_- j_2(x) \jt \jt \jt \rangle_{ff} +...  \big) \label{mix fourier intermediate 2}\\
+ \alpha \alpha_4 \frac{\tilde{\lambda}^2}{\tilde{N} (1+ \tilde{\lambda}^2)} \int d^3 x \big(\langle \partial_- \jt (x) \jt \rangle \langle j_2 (x) \jt \jt \jt \rangle_{cb}
- \frac{2}{5} \langle \jt(x) \jt \rangle \langle \partial_- j_2(x) \jt \jt \jt \rangle_{cb} +...  \big) \nonumber
\end{align}

%the five point function $\langle \partial \cdot j_4(x) \jt \jt \jt \jt \rangle$ factorizes into a product of a four point function times a two point function. 

%Four point functions $\sim  \frac{1}{\tilde{N}}$. The dependence on $\tilde{\lambda}$ and $\tilde{N}$ matches on both sides of (\ref{explicit}) and we obtain
%
%The rhs is
%\begin{align}
%\alpha_4 \alpha \frac{\tilde{\lambda}}{N(1+\lambda^2)} \Big( \int d^3 x  \langle \partial_{-} \jt(x) \jt(x_1) \rangle \big(  \langle j_2(x) \jt(x_2) \jt(x_3) \jt(x_4) \rangle_{ff}\\
%+\lambda \langle j_2(x) \jt(x_2) \jt(x_3) \jt(x_4) \rangle_{even} \big) \nonumber
%  - \frac{2}{5} \langle \jt(x) \jt(x_1) \rangle \nonumber \\
%   \big( \langle \partial j_2(x) \jt(x_2) \jt(x_3) \jt(x_4) \rangle_{ff}+\lambda \langle \partial j_2(x) \jt(x_2) \jt(x_3) \jt(x_4) \rangle_{even} \big)   + perm.  \Big) \nonumber 
%\end{align}
Let us equate (\ref{mix fourier intermediate 0}) and (\ref{mix fourier intermediate 2}). We see that the dependence on $\tilde{N}$ and $\tilde{\lambda}$ matches on both sides provided
\begin{align}
 \beta \langle (\partial_{-}\partial_{-} j_{- y} - \partial_{-}\partial_{y} j_{--}) \jt \jt \jt \rangle_{ff} + ... \label{mixed Fourier to solve 1} \\
  = - \alpha \alpha_4 \int d^3 x \big(\langle \partial_{-} \jt \jt \rangle  \langle j_2 \jt \jt \jt \rangle_{cb}  
  - \frac{2}{5} \langle \jt \jt \rangle \langle \partial j_2 \jt \jt \jt \rangle_{cb} + ... \big),   \nonumber 
\end{align}
\begin{align}
    \beta \langle (\partial_{-}\partial_{-} j_{- y} - \partial_{-}\partial_{y} j_{--}) \jt \jt \jt \rangle_{cb} + ...   \label{mixed Fourier to solve 2} \\
    =  \alpha \alpha_4 \int d^3 x \big(\langle \partial_{-} \jt \jt \rangle    \langle j_2 \jt \jt \jt \rangle_{ff} \nonumber 
  - \frac{2}{5} \langle \jt \jt \rangle  \langle \partial j_2 \jt \jt \jt \rangle_{ff} + ... \big).  \nonumber  
\end{align}

We solved (\ref{mixed Fourier to solve 1}) and (\ref{mixed Fourier to solve 2}) using a mixed Fourier transform. We define the mixed Fourier transform of a four point function $\langle \mathcal{O}_1(x_1) \mathcal{O}_2(x_2) \mathcal{O}_3(x_3) \mathcal{O}_4(x_4) \rangle$ as
\begin{align}
\langle \mathcal{O}_1(x_1) \mathcal{O}_2(x_2) \mathcal{O}_3(x_3) \mathcal{O}_4(x_4) \rangle \rightarrow
\int \frac{d^3 x_2 d^3 x_3}{(2 \pi i)^2} \langle \mathcal{O}_1(0) \mathcal{O}_2(x_2) \mathcal{O}_3(x_3) \mathcal{O}_4(\infty) \rangle e^{i (p_2 \cdot x_2 + p_3 \cdot x_3) } .
\end{align}
The advantage of the mixed Fourier transform with respect to a usual Fourier transform is that by placing an operator at the origin and another one at $\infty$ we take advantage of conformal symmetry. 

In mixed Fourier space we can get rid of the integrals in equations (\ref{mixed Fourier to solve 1}) and (\ref{mixed Fourier to solve 2}). For example, it is simple to see that the mixed Fourier transform of $\int d^3 x \langle \jt(x) \jt \rangle  \langle j_2(x) \jt \jt \jt \rangle$ is equal to
\begin{align}
\int d^3 x \langle \jt(x) \jt(x_1) \rangle  \langle j_2(x) \jt(x_2) \jt(x_3) \jt(x_4) \rangle \rightarrow \Big( \int d^3 x  \langle \jt(x) \jt(0) \rangle e^{i (p_2 + p_3) \cdot x} \Big) \\
\times \int \int d^3 x_2 d^3 x_3 e^{i \big( p_2 \cdot x_2 +  p_3 \cdot x_3 \big)}    \langle j_2(0) \jt(x_2) \jt(x_3) \jt(\infty) \rangle   \nonumber
\end{align}
which is just a product of mixed Fourier transforms. 

It turns out that $\langle j_2 \jt \jt \jt \rangle_{ff}$ is very simple in mixed Fourier space. Let us define $u_p = \frac{p_2^2}{p_1^2}, v_p = \frac{p_3^2}{p_1^2}$, where $p_1 = -p_2 - p_3$ . Then,
\begin{align}
\langle T_{\mu \nu}(0) \jt(p_2) \jt(p_3) \jt(\infty) \rangle_{ff}
  = \frac{f(u_p, v_p)}{p_1^4} \Big( (p_2)_{(\mu} \epsilon_{\nu ) \alpha \beta} (p_2)^{\alpha} (p_3)^{\beta} \Big) \\
   + \frac{f(v_p, u_p)}{p_1^4} \Big( (p_3)_{(\mu} \epsilon_{\nu ) \alpha \beta} (p_3)^{\alpha} (p_2)^{\beta} \Big) \nonumber,
\end{align}
where $f(u_p, v_p)= \frac{32}{3} \pi^2 (- \frac{1}{u_p}+\frac{1}{v_p} - \frac{1}{u_p v_p})$. Plugging this into (\ref{mixed Fourier to solve 1}) and (\ref{mixed Fourier to solve 2}) we obtain
\begin{align}
\langle T_{\mu \nu}(0) \jt(p_2) \jt(p_3) \jt(\infty) \rangle_{cb} 
= \frac{1}{|p_1|^3} \Big( (p_2)_{(\mu} (p_3)_{\nu)} - \frac{p_2 \cdot p_3}{3} \eta_{\mu \nu} \Big) f_1(u_p, v_p)  \\
 + \frac{1}{|p_1|^3} \Big( (p_2)_{\mu} (p_2)_{\nu} - \frac{p_2^2}{3} \eta_{\mu \nu} \Big) f_2(u_p, v_p) \nonumber 
 + \frac{1}{|p_1|^3} \Big( (p_3)_{\mu} (p_3)_{\nu} - \frac{p_3^2}{3} \eta_{\mu \nu} \Big) f_2(v_p, u_p) \nonumber, \nonumber
\end{align}
where
\begin{align}
f_1(u_p, v_p)= \frac{1}{2} \left(\frac{u_p}{v_p}+\frac{v_p}{u_p}\right)+\left(\frac{1}{u_p}+\frac{1}{v_p}\right)-\frac{3}{2 u_p v_p}, \\
f_2(u_p, v_p)=  \frac{u_p}{4 v_p}+\frac{v_p}{4 u_p}+\frac{1}{4 u_p v_p}+\frac{3}{2 u_p}-\frac{1}{2 v_p} \nonumber.
\end{align}
Finally, we can transform back to position space to get
\begin{align}
\langle T_{\mu \nu} \jt \jt \jt \rangle_{cb} = p  \times \Big( g_1(u, v) \big(V(1,2,3)_{\mu}V(1,2,3)_{\nu}- \frac{V(1,2,3)^2}{3}\eta_{\mu \nu}  \big) \\
 + g_2(u, v) \big(V(1,2,3)_{(\mu}V(1,3,4)_{\nu)}- \frac{V(1,2,3)\cdot V(1,3,4)}{3}\eta_{\mu \nu}  \big) \nonumber \\
 + g_3(u, v) \big(V(1,3,4)_{\mu}V(1,3,4)_{\nu}- \frac{V(1,3,4)^2}{3}\eta_{\mu \nu}  \big) \nonumber\Big), \nonumber
\end{align}
where $p= \frac{1}{(x_{12} x_{13} x_{14})^{\frac{10}{3}} (x_{23} x_{24} x_{34})^{\frac{1}{3}} }$, $V(i; j, k) = \frac{x_{ij}^2 (x_{ik})_{\mu}-x_{ik}^2 (x_{ij})_{\mu}}{x_{jk}^2}$ and
\begin{align}
g_1(u, v)= \frac{u^{2/3} v^{2/3}}{4 \pi ^3}-\frac{v^{2/3}}{4 \pi ^3 u^{4/3}}+\frac{v^{5/3}}{2 \pi ^3 u^{4/3}}-\frac{v^{8/3}}{4 \pi ^3 u^{4/3}}+\frac{v^{2/3}}{2 \pi ^3 \sqrt[3]{u}}+\frac{v^{5/3}}{2 \pi ^3 \sqrt[3]{u}}, \\
g_2(u, v)=\frac{u^{2/3} v^{2/3}}{2 \pi ^3}+\frac{u^{2/3}}{2 \pi ^3 \sqrt[3]{v}}+\frac{u^{5/3}}{4 \pi ^3 \sqrt[3]{v}}+\frac{v^{2/3}}{2 \pi ^3 \sqrt[3]{u}}+\frac{v^{5/3}}{4 \pi ^3 \sqrt[3]{u}}-\frac{3}{4 \pi ^3 \sqrt[3]{u} \sqrt[3]{v}},  \nonumber \\
g_3(u, v)= \frac{u^{2/3} v^{2/3}}{4 \pi ^3}-\frac{u^{2/3}}{4 \pi ^3 v^{4/3}}+\frac{u^{5/3}}{2 \pi ^3 v^{4/3}}-\frac{u^{8/3}}{4 \pi ^3 v^{4/3}}+\frac{u^{2/3}}{2 \pi ^3 \sqrt[3]{v}}+\frac{u^{5/3}}{2 \pi ^3 \sqrt[3]{v}}. \nonumber 
\end{align}
The result agrees with \cite{Li:2019twz}. For correlators of type $\langle j_s \jt \jt \jt \rangle$ with $s \geq 4$, the mixed Fourier transform is not so simple, so in practice it was not useful.

\section{Miscellaneous formulas}\label{app:dump}

In this appendix we write some formulas we used in the text. The nonzero coefficients in equation (\ref{pseudoconservation polynomial}) are
\begin{align}\label{pol pseudo coefs}
a_{1,-1,-1}= -(\gamma_{14}-1) \left(2 \gamma_{14}^2-\gamma_{14} (4 k+5)+2 k^2+5 k+2\right) \left(k^2-2 k s-k+s^2+s\right), \\
 a_{0,0,0}=-\frac{1}{2} \left(2 \gamma_{14}^2-\gamma_{14} (4 k+5)+2 k^2+5 k+2\right)
 (-2 \gamma_{12} (k+s)+\gamma_{14} (2 k-2 s+1)+s (2 s+1)) \nonumber \\
 \times  (k-s), ~~~  a_{1,-1,0}= -\frac{1}{2}  (2 \gamma_{12}^2+\gamma_{12} (4 k-4 s-1)+2 k^2-k (4 s+1)+2 s^2+s-1) (\gamma_{14}-1) \nonumber \\
 \times \left(2 k^2-4 k s+k+s (2 s-1)\right) , ~~~ \nonumber
a_{0,-1,0}= \frac{1}{2} (\gamma_{14}-1) \left(2 k^2-4 k s+k+s (2 s-1)\right) \\
\times ( 2 \gamma_{12}^2+\gamma_{12} (4 \gamma_{14}-4 s-3)+2 \gamma_{14}^2-\gamma_{14} (4 s+3)+s (2 s+3)  ), \nonumber  \\
  a_{-1,0,1}=-\frac{1}{2} (\gamma_{12}-1) \left(2 k^2+3 k+1\right) ( 2 \gamma_{12}^2+\gamma_{12} (4 \gamma_{14}-4 s-3)+2 \gamma_{14}^2-\gamma_{14} (4 s+3)+s (2 s+3) ) \nonumber \\
    a_{-1,1,1} = \frac{1}{2} (\gamma_{12}-1) \left(2 k^2+3 k+1\right) \left(2 \gamma_{14}^2-\gamma_{14} (4 k+5)+2 k^2+5 k+2\right) \nonumber \\
     a_{0, 0,1} = \frac{1}{2} (k+1) \left(2 \gamma_{12}^2+\gamma_{12} (4 k-4 s-1)+2 k^2-k (4 s+1)+2 s^2+s-1\right) \nonumber \\ 
   \times (2 \gamma_{12} k+\gamma_{12}-2 \gamma_{14} (k-2 s+1)-s (2 s+1)) \nonumber , ~~~ a_{-1, 1 , 2} = (\gamma_{12}-1) \left(k^2+3 k+2\right)  \\
   \times (2 \gamma_{12}^2+\gamma_{12} (4 k-4 s-1)+2 k^2-k (4 s+1)+2 s^2+s-1 ) . \nonumber
\end{align}

The position space correlator $\langle j_4 \jt \jt \jt \rangle_{cb}$ is given by (\ref{corr with struct}), with
\begin{align}\label{spin 4 position}
(1 + \sqrt{u} + \sqrt{v})^3 \times p \times f_0(u, v) = -\frac{\left(\sqrt{v}-1\right)^2 \left(\sqrt{v}+1\right)^4}{10 u^{9/2} v}+\frac{\left(17 v-\sqrt{v}+17\right) \left(\sqrt{v}+1\right)^2}{60 u^{7/2} v}\\
+\frac{3 v+2 \sqrt{v}+3}{30 u^{5/2} v} +\frac{1}{90 u^{3/2} v} -\frac{\left(\sqrt{v}-1\right)^2 \left(\sqrt{v}+1\right)^5}{30 u^5 v}+\frac{\left(v+6 \sqrt{v}+1\right) \left(\sqrt{v}+1\right)^3}{30 u^4 v} \nonumber \\
+\frac{47 v^{3/2}+42 v+42 \sqrt{v}+47}{180 u^3 v}+\frac{\sqrt{v}+1}{30 u^2 v}, \nonumber \\
(1 + \sqrt{u} + \sqrt{v})^3 \times p \times f_1(u, v) = \frac{7 v+6 \sqrt{v}+11}{45 u^{3/2} v^2} \nonumber \\
+\frac{\left(3 v^{3/2}-3 v+7 \sqrt{v}-7\right) \left(\sqrt{v}+1\right)^3}{10 u^{7/2} v^2}+\frac{22 v^{3/2}+17 v^2+32 v+44 \sqrt{v}+17}{30 u^{5/2} v^2} \nonumber \\
 +\frac{\left(3 v^{3/2}-3 v+7 \sqrt{v}-7\right) \left(\sqrt{v}+1\right)^4}{30 u^4 v^2}+\frac{\left(47 v^{3/2}-7 v+91 \sqrt{v}-35\right) \left(\sqrt{v}+1\right)^2}{90 u^3 v^2} \nonumber \\
+\frac{16 v^{3/2}+15 v+27 \sqrt{v}+31}{45 u^2 v^2}+\frac{\sqrt{v}+1}{15 u v^2}+\frac{1}{45 \sqrt{u} v^2}, \nonumber \\
(1 + \sqrt{u} + \sqrt{v})^3 \times p \times f_2(u, v) = \frac{\left(10 v^2+28 v-35\right) \left(\sqrt{v}+1\right)^2}{30 u^{5/2} v^3} \nonumber \\
+\frac{69 v^{3/2}+49 v^2+172 v+168 \sqrt{v}+49}{90 u^{3/2} v^3}+\frac{\left(10 v^2+28 v-35\right) \left(\sqrt{v}+1\right)^3}{90 u^3 v^3} \nonumber \\
+\frac{172 v^{3/2}+46 v^{5/2}+78 v^2+196 v-21 \sqrt{v}-77}{90 u^2 v^3}+\frac{23 v+30 \sqrt{v}+29}{45 \sqrt{u} v^3} \nonumber \\
+\frac{\sqrt{u}}{9 v^3}+\frac{49 v^{3/2}+78 v+114 \sqrt{v}+94}{90 u v^3}+\frac{\sqrt{v}+1}{3 v^3}, \nonumber \\
f_3(u,v) = f_1(v, u), ~~~ f_4(u, v) = f_0 (v,u), \nonumber
\end{align}
where $p = |x_1-x_3|^{-18}  |x_2-x_3|^{7} |x_2-x_4|^{-11}  |x_3-x_4|^{7} $.

\section{Contact interactions for scattering amplitudes}\label{app:scattering}

In section \ref{bound chaos section} we analyse contact interactions in AdS. We conclude that AdS contact diagrams for $\langle j_s \jt \jt \jt \rangle$ violate the bound on chaos if $s \geq 4$, whereas for $s=2$ there is one parity even and one parity odd contact term that does not violate the bound on chaos.

In this appendix we consider the same problem at the level of $4$ dimensional scattering amplitudes. We consider on-shell $2 \rightarrow 2$ scattering amplitudes of the type $\langle \Phi_s \phi \phi \phi \rangle$, where the particle $\Phi_s$ at position $1$ is massless and has spin $s$ and the three other particles are identical massive scalars $\phi$ of mass $m$. It is expected that there is a bijection between contact terms for $d$-dimensional CFT's and $d+1$ dimensional flat space scattering amplitudes \cite{Costa:2011mg}.

For scattering amplitudes the analog of the bound on chaos is the statement that the amplitude cannot grow more than quadratically in the Regge limit \cite{Caron-Huot:2021rmr, Chowdhury:2019kaq}. This is the \textit{Classical Regge Growth} (CRG) conjecture of \cite{Chowdhury:2019kaq} which states that \textit{the S- Matrix of a consistent classical theory never grows faster than $s^2$ at fixed t - at all physical values of momenta and for every possible choice of the normalized polarization vector $\zeta_i$}. 

In this appendix we perform some calculations for contact scattering amplitudes that support the results of section (\ref{bound chaos section}). We construct contact scattering amplitudes for $\langle \Phi_2 \phi \phi \phi \rangle$ and conclude that there is only one parity even and only one parity odd contact scattering amplitude compatible with the CRG conjecture\footnote{A similar analysis for the case of four scalars was done in \cite{Turiaci:2018dht}. There it was found that there are three contact diagrams that do not violate the Regge bound, whose scattering amplitudes are given by $1, s \times t \times u$ and $s^2 + t^2 + u^2$ respectively.}. 

%Furthermore, we examine contact scattering amplitudes for $\langle \Phi_s \phi \phi \phi \rangle$ with $s \geq 4$ and conclude that they all diverge faster than quadratically in the Regge limit.

Let us discuss the parity even case. In that case the scattering amplitude $T(\zeta, p_1, p_2, p_3)$ is a function of the polarization $\zeta$ and the momenta. Notice we used momentum conservation to eliminate $p_4$. Furthermore the polarization obeys $\zeta \cdot \zeta = p_1 \cdot \zeta = 0$. Thus we have
\begin{align}
T(\zeta, p_1, p_2, p_3) = (\zeta \cdot p_2 )^2 f_1(s, t) + (\zeta \cdot p_3 )^2 f_2(s, t)  + (\zeta \cdot p_2 ) (\zeta \cdot p_3 ) f_3(s, t),
\end{align}
where we define the Mandelstam invariants as 
\begin{align}
s= - (p_1 + p_2)^2, ~~~ t= - (p_1 + p_3)^2, ~~~ u= - (p_1 + p_4)^2 .
\end{align}
The amplitude is constrained by crossing symmetry and gauge invariance. Gauge invariance is the statement that the amplitude is invariant under $\zeta \rightarrow \zeta  + \lambda_1 p_1$, where $\lambda_1$ is an arbitrary real number. Crossing symmetry and gauge invariance lead to the constraints
\begin{align}
f_2(s, t ) = \frac{(m^2 - s)^2}{(m^2 - t)^2} f_1 (s, t), ~~ f_3(s, t ) = \frac{2 (s - m^2)}{(m^2 - t)} f_1 (s, t)\\
 f_1 (s, t) = \frac{(m^2 - t)^2}{(m^2 - s)^2} f_1 (t, s), ~~ f_1(s, t) = f_1 (u, t) .
\end{align}

We solved the above expressions using polynomials $f_i (s, t) = \sum c^i_{n_1, n_2, n_3} s^{n_1} t^{n_2} m^{n_3}$. The solution with polynomials of lowest degree is 
\begin{align}\label{S Matrix spin 2 even}
T= (\zeta \cdot p_2)^2  (p_1 \cdot p_3)^2 + (\zeta \cdot p_3)^2  (p_1 \cdot p_2)^2 -2 (\zeta \cdot p_2)(\zeta \cdot p_3) (p_1 \cdot p_2)(p_1 \cdot p_3) .
\end{align}

Let us analyse the Regge limit of this expression in light of the CRG conjecture. We need to parametrize $\zeta$, which we do in the following manner \cite{Chowdhury:2019kaq}
\begin{align}\label{parametrization}
\zeta = \zeta^{\perp} +  \zeta^{\parallel}, ~~~ \zeta^{\parallel} = \alpha_1 \sqrt{\frac{s t }{u}} (\frac{p_2}{s-m^2} - \frac{p_3}{t-m^2}) + a_1 p_1.
\end{align}
$\zeta^{\perp}$ is the component of $\zeta$ that is perpendicular to the plane generated by $p_1$, $p_2$ and $p_3$, $\zeta^{\parallel}$ is the parallel component. Let us explain the logic for this parametrization. A priori, $\zeta$ depends on $4$ independent parameters, but due to the conditions $\zeta^2 = \zeta \cdot p_1 =0$ it only depends on two, which we call $\alpha_1$ and $a_1$. However, the component $a_1$ is not physical due to the gauge symmetry of the S-Matrix. 

We can now take the expression (\ref{parametrization}) for $\zeta$, plug it into (\ref{S Matrix spin 2 even}) and consider the limit $s \rightarrow \infty$ with $t$ fixed. We find that the amplitude grows like $s^2$. Thus, it obeys the CRG conjecture.

%Let us analyse the Regge limit of this expression. The analog of MSS coordinates (\ref{MSS coordinates}) for momentum space is to take 
%\begin{align}
%p_1 \cdot p_2 \rightarrow p_1 \cdot p_2, ~~ p_1 \cdot p_3 \rightarrow \lambda p_1 \cdot p_3, ~~ p_1 \cdot p_4 \rightarrow \lambda p_1 \cdot p_4, ~~ p_2 \cdot p_3 \rightarrow \lambda p_2 \cdot p_3\\
%p_2 \cdot p_4 \rightarrow \lambda p_2 \cdot p_4,~~ p_3 \cdot p_4 \rightarrow  p_3 \cdot p_4,~~ \zeta \cdot p_2 \rightarrow  \zeta \cdot p_2,~~ \zeta \cdot p_3 \rightarrow \lambda \zeta \cdot p_3. \nonumber
%\end{align}
%CHECK IF THERE ARE NO SUBTLETIES HERE!!!!
%and consider the limit $\lambda \rightarrow \infty$. The scattering amplitude should be bounded by $\lambda^2$. In the expression above $T \sim \lambda^2$, so the amplitude respects this bound. For contact diagrams with higher powers of $s$ and $t$ we checked that they grow faster than $\lambda^2$ and for this reason they are not allowed.

A similar analysis can be performed for the parity odd contact term. The term that is Regge bounded is
\begin{align}
T = \epsilon_{\mu_1 \mu_2 \mu_3 \mu_4} \zeta^{\mu_1} p_1^{\mu_2} p_2^{\mu_3} p_3^{\mu_4} \Big( ( p_1 \cdot p_3 )(\zeta \cdot p_2) -  ( p_1 \cdot p_2 )(\zeta \cdot p_3) \Big).
\end{align}

\bibliography{bibforproposal}
\bibliographystyle{ieeetr}

%\bibliography{Asymptotics_and_Borel_Summability,S0377042714003161,autobib}

\end{document}